\documentclass[aps,prb,epsfig,amsfonts,twocolumn,showpacs,floats,goodfloats]{revtex4}

\usepackage{epsf}
\usepackage{graphics}
\usepackage{bbm}

\newcommand{\myscalebox}[1]{\scalebox{0.42}[0.42]{#1}}

\newcommand{\bc}{\begin{center}}
\newcommand{\ec}{\end{center}}
\newcommand{\be}{\begin{equation}}
\newcommand{\ee}{\end{equation}}
\newcommand{\la}{\langle}
\newcommand{\ra}{\rangle}
\newcommand{\beqn}{\begin{eqnarray}}
\newcommand{\eeqn}{\end{eqnarray}}

\newcommand{\ql}{q_{l}}
\newcommand{\ml}{\mu_{l}}

\begin{document}


\title{Low Energy Excitations in Spin Glasses from Exact Ground 
States}

\author{Matteo Palassini}
\address{University of California, 3333 California Street, Suite 415, 
San Francisco, CA 94118}

\author{Frauke Liers and Michael Juenger}
\address{Institut f\"ur Informatik, Universit\"at zu K\"oln, D50969 Cologne,
Germany}

\author{A. P. Young}
\address{Physics Department, University of California, Santa Cruz CA 95064}

\date{\today}


\begin{abstract}
We investigate the nature of the low-energy, large-scale excitations
in the three-dimensional Edwards-Anderson Ising spin glass with 
Gaussian couplings and free boundary conditions, by studying the 
response of the ground state to a coupling-dependent perturbation 
introduced previously.
The ground states are determined {\em exactly} for system sizes up
to $12^3$ spins using a branch and cut algorithm.
The data are consistent with a picture where the surface 
of the excitations is not space-filling, such
as the droplet or the ``TNT'' picture,  with only minimal corrections
to scaling. When allowing for  very large 
corrections to scaling, the data are also consistent with 
a picture with space-filling surfaces,
such as replica symmetry breaking. The energy
of the excitations scales with their size with a small
exponent $\theta'$, which is compatible with zero if we allow
moderate corrections to scaling.
We compare the results with data for periodic boundary conditions 
obtained with a genetic algorithm, and discuss the effects of different 
boundary conditions on corrections to scaling.  Finally, we analyze the 
performance of our branch and cut algorithm, 
finding that it is correlated with the existence of large-scale, 
low-energy excitations.

\end{abstract}

\maketitle

\section{Introduction}
\label{sec:intro}

There is still considerable debate about the nature of the spin glass state
in finite dimensional spin glasses. 
Two principal theories have been investigated: the
``droplet theory'' proposed by Fisher and Huse\cite{fh}
(see also Refs.~\onlinecite{bm,mcmillan}),
and the replica symmetry
breaking picture of Parisi\cite{parisi,mpv,bindery}.
In the droplet theory the lowest energy excitation of length
scale $\l$ (a ``droplet'') has energy of order $\l^\theta$ where
$\theta\, $ is a positive exponent. 
Furthermore, the droplets have a surface with fractal dimension, $d_s$,
less than the space dimension $d$.

Replica symmetry breaking (RSB) is well established in mean field theory,
but it remains to be proven in finite dimensions. 
The precise nature of RSB
in finite dimensions is not uniquely defined
but it is generally agreed that a key feature 
of RSB is the existence of excitations whose energy, unlike that 
of droplets, remains of order unity even as their size tends to infinity. 
Furthermore, upon the creation of 
such a large scale,  finite energy  excitation,  a finite 
fraction of the bonds change 
state (from satisfied to unsatisfied, or vice-versa) or, equivalently,
the surface of these excitations 
is space filling, {\em i.e.} $d_s = d$.

Recently, Krzakala and Martin\cite{km} (KM),
and two of us\cite{py-bulk} (PY),
have argued, on the basis of numerical calculations at zero temperature,
in favor of an intermediate scenario where there are
large scale excitations whose energy does not increase with
size, as in RSB, but which have a surface
with $d_s < d$. Following KM we shall denote this the ``TNT'' scenario.
In the TNT scenario  it is necessary to introduce
{\em two} exponents which describe the growth of the energy of an excitation
of scale $L$: (i) $\theta \ (> 0)$ such that $L^\theta$ is the typical change
in energy when the boundary conditions are changed, for example from periodic
to anti-periodic, in a system of size $L$, and
(ii) $\theta^\prime$, which characterizes the energy of
clusters excited within the system for a {\em fixed}\/ set of boundary
conditions ($\theta'$ was called $\theta_g$ in Ref.\onlinecite{km}). 
The TNT picture has been
challenged (although in opposite senses) by
Marinari and Parisi\cite{mapabulk} and by Middleton\cite{middleton}. 
Subsequently, low temperature  Monte Carlo simulations\cite{kpy} have found
results consistent with the TNT scenario.
The RSB, droplet, TNT and some other scenarios have been also studied by 
Newman and Stein\cite{ns-old,ns}. For some recent related work, 
see Refs.~\onlinecite{mezard,moore}.

The work of KM and PY determined the ground state with and without a certain
perturbation (which was different in the two cases), designed so that the
ground state of the perturbed system is a large scale excitation of the
original system.  They used \textit{heuristic} algorithms,
i.e.  algorithms which are
not {\em guaranteed}\/ to give the exact ground state, although both KM and PY
argue that they do find the exact ground state in most cases.

In this paper,  we reconsider the problem of determining $\theta'$ and
$d_s$, following 
the perturbation approach of PY, 
described in Section \ref{sec:method}, but we
apply an \textit{exact} algorithm, known as ``branch and cut''
\cite{junger-original}, so we are guaranteed that the true ground state 
is reached every time.
Exact optimization algorithms have been used before 
for spin glasses, see e.g. Refs.~\onlinecite{kobe,dombgreen,desim},
but, to our knowledge, their use in three-dimensions
has been restricted to smaller sizes than studied here, and they
were not used to investigate the real-space structure of the 
low-energy excitations.

Our implementation of the branch and cut technique can handle
significantly larger sizes for free
boundary conditions (bc) than for periodic bc\cite{liers}, so we
use free bc here. We consider a different (and enlarged) set of 
observables than PY, in the attempt to gain a fuller understanding
of what picture fits better the whole set of observables. 
We also perform a similar analysis 
of the data of PY, who used periodic bc, in order to 
investigate the effects of 
different types of boundary conditions.
 The various pictures discussed refer to the large volume limit, 
while the sizes that can be currently reached are rather small. 
We will therefore pay particular attention to properly take into account 
{\em corrections to scaling}. In particular, we will try to determine what
values of the parameters $\theta'$ and $d-d_s$ fit the data
in the more ``natural'' way,  namely with 
 the smallest corrections to scaling for the range of sizes considered.

 A summary of our results is as follows.
We find that for {\em periodic}\/
bc, a simple scaling ansatz fits the results in a
natural way, {\em i.e.}\/ with negligible corrections to scaling and no
adjustable parameters besides $d-d_s$ and $\theta'$. This
gives $d-d_s=0.42 \pm
0.03$; $\theta'=-0.01\pm 0.03$ (the meaning of the error bars will be 
explained later), which agrees with the results of PY, and is
compatible with the TNT picture. We cannot rule out crossover to either the
droplet or the RSB picture at length scales larger than our system sizes, but
these scenarios,  especially the latter, 
would require larger corrections to scaling than the TNT
picture.

For {\em free}\/ bc, {\em all}\/ forms of fitting require some corrections to scaling.
 The most natural scenario, in the sense explained
above, gives $d-d_s=0.45 \pm 0.02$; $\theta'=0.18\pm 0.03$, with 
small corrections (of the order of 3$\%$), which 
is compatible with the droplet picture. Allowing somewhat larger 
corrections (of order 10$\%$), the data are also compatible with $\theta'=0$,
namely with the TNT picture. Finally, if we allow for 
very large corrections, the data are also consistent with the RSB picture.

In the second part of the paper, we analyze the performance of the
branch and cut algorithm. We find that the number of
elementary operations required to find the ground state
increases exponentially with the size, as expected since computing 
a ground state of a
three-dimensional spin glass system is an
$\mathcal{NP}$-hard problem\cite{barahona}. 
We also find, interestingly enough,
 that the CPU time is larger for samples in which
there is an excited state close in energy
to the ground state energy, yet different from the ground state in the
orientation of a large number of spins. We are not aware of any previous
quantitative measures of this trend,  which we expect to be common to other algorithms as well.

The rest of this paper is organized as follows.
In Section \ref{sec:method} we describe the method of perturbing the ground
states to get information about low energy excitations, introduced by PY. 
Our results for the nature of the large scale, low energy
excitations are given in Section~\ref{sec:results}. 
A short description of the branch and cut algorithm used is given
in Section \ref{sec:b&c}, 
and the performance of the algorithm is analyzed in  Section~\ref{sec:performance}.
Our conclusions are summarized in Section~\ref{sec:discussion}.

\section{Ground state perturbation method}
\label{sec:method}

The Hamiltonian of the spin glass model is given by
\begin{equation}
{\cal H} = -\sum_{\langle i,j \rangle} J_{ij} S_i S_j ,
\label{eq:ham}
\end{equation}
where the sites $i$ lie on a 
simple cubic lattice with $N=L^3$ spins in dimension $d=3$, $S_i=\pm 1$,
and the $J_{ij}$ are nearest-neighbor interactions chosen from a
Gaussian distribution with zero mean and standard deviation unity. Free
boundary conditions are applied in all directions.

For a given set of bonds we
determine the exact ground state using a branch and cut algorithm
discussed in Section ~\ref{sec:b&c}.
Let $S_i^{(0)}$ be the ground state spin configuration. As in PY
we then perturb the couplings $J_{ij}$ by an amount
proportional to $S_i^{(0)} S_j^{(0)}$ in order to increase the
energy of the ground state relative to the other states and therefore 
possibly induce a change in the ground state.
This perturbation, which depends upon a positive parameter $\epsilon$, 
is defined by
\begin{equation}
\Delta {\cal H}_\epsilon  =  {\epsilon \over N_b} \sum_{\langle i,j
\rangle}
S_i^{(0)} S_j^{(0)} S_i S_j,
\end{equation}
where
$N_b\, =dL^{d-1}(L-1)$ is the number of bonds in the Hamiltonian.
We denote the unperturbed ground state energy by $E^{(0)}$ and the perturbed
energy of the \textit{same} state by ${E}_\epsilon^{(0)}$.
The energy of the unperturbed
ground state will thus increase exactly by an amount
$ \Delta E^{(0)} \equiv {E}_\epsilon^{(0)} - E^{(0)} = \epsilon .$
The energy of any other state, $\alpha$ say, will increase by the lesser
am\-ount
$ \Delta E^{(\alpha)} \equiv {E}_\epsilon^{(\alpha)} - E^{(\alpha)}
= \epsilon\ \ql^{(0, \alpha)},$
where $\ql^{(0, \alpha)}$ is the ``link overlap'' between the states
``0'' and $\alpha$, defined by
\begin{equation}
\ql^{(0, \alpha)} = {1 \over N_b}\sum_{\langle i,j \rangle} S_i^{(0)}
S_j^{(0)} 
S_i^{(\alpha)} S_j^{(\alpha)} ,
\end{equation}
in which the sum is over all the $N_b$ nearest neighbor bonds. 
Note that the {\em total} energy of the states changes by an
amount of order unity.

As we apply the perturbation, the energy {\em difference} between a low energy 
excited state and the ground state decreases by the amount
\begin{equation}
\Delta  E^{(0)} - \Delta E^{(\alpha)} = 
\epsilon \ (1 - \ql^{(0, \alpha)}) .
\label{de}
\end{equation}
If there is at least one excited state such that $E^{(\alpha)} - E^{(0)} <
\Delta  E^{(0)} - \Delta E^{(\alpha)}$, then 
one of these excited states will become the ground state of the perturbed
Hamiltonian. We denote the new ground state spin
configuration by $ \tilde{S}_i^{(0)}$, and indicate by
$\ql$ and $q$, with no indices, the 
link- and spin-overlap
between the new and old ground states ${S}_i^{(0)}$ and $ \tilde{S}_i^{(0)}$,
where $q$ is defined as usual by $q=1/N \sum S^{(0)}_i \tilde{S}^{(0)}_i$.

Due to the spin flip symmetry of the Hamiltonian (\ref{eq:ham}), the ground
state is doubly degenerate, and therefore the distribution of $q$ is
symmetric \cite{reality} around $q=0$.
Hence, in the rest of the paper we will restrict
ourselves to $q\ge 0$ without loss of information.

Consider  the 
probability $P(\epsilon,L)$
(with respect to the random couplings) 
that $q$ is less than unity, {\em i.e.}\/ 
that ${S}_i^{(0)}$ and 
$\tilde{S}_i^{(0)}$ differ in a finite fraction of the spins.
As discussed by PY, we  assume that
$P(\epsilon,L)$ is dominated by those
samples in which ${S}_i^{(0)}$ and $ \tilde{S}_i^{(0)}$ differ by 
flipping a single 
connected cluster of spins, with linear size $L$. Deviations from this 
assumption give rise to corrections to scaling, as pointed out 
by Middleton\cite{middleton}, and will be analyzed in Section \ref{sec:results}. 
There are two energy scales in the problem: 
the typical energy above the ground state of such an excitation,
which scales as $L^{\theta'}$ ($\theta'=\theta$ in the droplet picture),
and the threshold energy of Eq.~(\ref{de}), which scales as 
$\epsilon L^{-(d-d_s)}$ since  $1-\ql$ is proportional to 
the surface of the excitation, $1-\ql \sim L^{-(d-d_s)}$.  
Hence, the dimensionless probability $P(\epsilon,L)$ is a function of the ratio of
these two energy scales:
\begin{equation}
P(\epsilon,L) = g(\epsilon L^{-\mu}) \,,
\label{scalingP2}
\end{equation}
where $g(x)$ is a scaling function and
\begin{equation}
\mu \equiv \theta' + d - d_s \,.
\label{mudef}
\end{equation}
From this we obtain scaling relations for various observables.
For example, since $1-q\sim O(1)$ and $1-\ql \sim L^{-(d-d_s)}$,
we obtain\cite{py-bulk}:
\begin{eqnarray}
\la 1 -  q\ra & = &  F_q(\epsilon L^{-\mu})
\label{scalingq} \\
\la  1 -  \ql \ra & = &  L^{-(d-d_s)} F_{q_{l}}(\epsilon L^{-\mu}) ,
\label{scaling}
\end{eqnarray}
where  $\langle \cdots \rangle$ is the
average with respect to the random couplings.
By measuring $\la 1 -  q\ra$ and $\la 1 - \ql\ra$
we can then determine $d-d_s$ and $\theta'$,  the two
exponents discriminating the various pictures of the spin glass phase
discussed in  Section~\ref{sec:intro}.

For small $\epsilon$, we expect the probability that the ground state
changes to be proportional to $\epsilon$ (for fixed $L$), which implies 
$g(x)\sim x$ for $x \to 0$.
Hence $F_q(x)$ and $F_{q_{l}}(x)$ also vary linearly for small $x$, and
the {\em asymptotic scaling}\/ behavior for $ L \gg \epsilon^{1/\mu}$ is
\begin{eqnarray}
\langle 1 -  q\rangle  &\sim& \epsilon L^{-\mu} , \label{1mq}  \\
\langle 1 - \ql \rangle &\sim& \epsilon  L^{-\ml} \label{1mql} ,
\end{eqnarray}
where
\begin{equation}
\ml \equiv \theta' + 2(d - d_s) \, .
\label{muldef}
\end{equation}

In the RSB case, $d-d_s=\theta'=0$, and therefore $\mu=\mu_l=0$. 
The scaling relations in Eqs. (\ref{scalingq}, \ref{scaling}) reduce in this case to
\begin{equation}
\langle 1-q \rangle = F_q(\epsilon)\, , \quad 
\langle 1-\ql \rangle = F_{\ql}(\epsilon) \quad \quad {\mbox{(RSB)}} \, ,
\label{scalingRSB} 
\end{equation}
and the asymptotic scaling behavior to
\begin{equation}
\langle 1 -  q\rangle \sim \epsilon,\, \, \, \,
\langle 1 - \ql \rangle 
\sim \epsilon \quad 
\mbox{(RSB)} \, .
\label{1mqRSB}  
\end{equation}

We see that both scaling and asymptotic scaling are in a sense 
trivial in RSB since the $L$ dependence disappears. Nevertheless, we will
still use the term {\em scaling}.

It is also convenient to analyze just those samples in which the unperturbed
and perturbed ground states are very different, i.e. where
$q \le q_{\max}$, a threshold value. Denoting such restricted averages by
$\langle \cdots \rangle_c$, we have
\begin{equation}
\la 1 - \ql \rangle_c =   L^{-(d-d_s)} F^c_{\ql}(\epsilon L^{-\mu}) .
\label{eqscalqlim}
\end{equation}
This is of the same form as in
Eq.~(\ref{scaling}), but, for sufficiently small $q_{\max}$,
the behavior of the scaling functions $F_{\ql}(x)$ and
$F^c_{\ql}(x)$ at small argument will be different for the following reason.
If we average over all samples we need to include the probability
$P(\epsilon,L)$ that the perturbation generates
an excitation with $q<1$. This is proportional to $\epsilon L^{-\mu}$
for $\epsilon L^{-\mu} \ll 1$, which 
is the reason why $F_q(x) \sim x$ for small
$x$. However, this factor is automatically taken into account in the {\em
selection}\/ of the samples in the restricted average in Eq.~(\ref{eqscalqlim}),
and so should not be
included again when performing the average. As a result, 
$F^c_{\ql}(x)$ tends to a constant for $x \to 0$, therefore
the asymptotic scaling is
\begin{equation}
\la 1 -  \ql \rangle_c \sim L^{-(d-d_s)}  .
\label{1mqlp}
\end{equation}
In particular, in RSB this becomes 
\begin{equation}
\la 1 -  \ql \rangle_c \sim  \mbox{const.} \quad \mbox{(RSB)} \,.
\label{1mqlpRSB} 
\end{equation}
Note that in both cases the asymptotic scaling is {\em independent} of $\epsilon$.

When analyzing the numerical data, we must be aware that there are
corrections to both (simple) scaling and asymptotic scaling that occur when
$L$ is not large enough.
Corrections to simple scaling take the form of {\em additive}\/ corrections
to relations such as
Eqs.~(\ref{scalingP2}), (\ref{scalingq}), (\ref{scaling}), and
(\ref{eqscalqlim}), whose amplitude
is characterized by a correction-to-scaling exponent $\omega$. For example,
including the leading correction,
Eq.~(\ref{eqscalqlim}) becomes
\begin{equation}
\la  1 -  \ql \rangle_c = {1 \over L^{d-d_s}} \left\{
F^c_{q_{l}}(\epsilon L^{-\mu})
 + {1 \over L^{\omega}}
G_{q_{l}}(\epsilon L^{-\mu}) \right\} .
\label{correction}
\end{equation}
For $\epsilon L^{-\mu} \to 0$, this gives the correction
to asymptotic scaling corresponding to Eq.~(\ref{1mqlp})
\begin{equation}
\la 1 -  \ql \rangle_c = {1 \over
L^{d-d_s}} \left(a+ {b \over  L^{\omega}}\right).
\end{equation}
For the RSB case, this goes over to 
\begin{equation}
\la 1 -  \ql \rangle_c = a + {b \over L^\omega} ,
\label{rsbcorr}
\end{equation}
rather than Eq.~(\ref{1mqlpRSB}).

Even when these corrections to (simple) scaling are negligible 
and the scaling {\em form}\/, such
as Eq.~(\ref{eqscalqlim}), is valid, 
the argument of the scaling function
may not be sufficiently small for asymptotic scaling to hold.
In this regime, when fitting the data to asymptotic scaling we have to
consider {\em further} corrections to (asymptotic) scaling, 
whose form is obtained by expanding the scaling function in its argument.
For example, the leading correction to Eq.~(\ref{1mqlp}), coming from
expanding the $F^c_{\ql}$ in Eq.~(\ref{eqscalqlim}) to second order, will be
\begin{equation}
\la 1 -  \ql \rangle_c =
{1 \over  L^{d-d_s}} \left(a+ b {\epsilon \over L^\mu}\right)
\end{equation}
which goes over to $\la 1 -  \ql \rangle_c = a + b \, \epsilon$ in RSB.
In general, both types of corrections need to be borne in mind
when fitting the data.

\section{Results}
\label{sec:results}

We applied the perturbation method described in the previous Section to
systems of size $L=4,6,8,10,$ and $12$.
For each size, we considered five values of the perturbation strength
$\epsilon$, namely $\epsilon/\tau = {1\over 4}, {1\over 2}, 1, 2,$ and 4,
where $\tau = \sqrt{6}$ is the mean field transition temperature,
except for $L=12$
for which only  $\epsilon/\tau = {1\over 4}$ and $1$ were considered. We choose
this value of $\tau$ so we can compare our results with the results of PY for
periodic bc.  In order to discriminate between the different pictures, it is
important to have high statistics.  Table \ref{tab_samples} reports the
number of samples computed for each size.
Note that the number of samples necessary to achieve a given
statistical error increases as $\epsilon$ decreases, since the fraction of
samples in which $\tilde{S}^{(0)}\neq S^{(0)}$ decreases.

\begin{table}
\begin{center}
\begin{tabular}{r@{\hspace{0.3cm}}r@{\hspace{0.4cm}}r@{\hspace{0.4cm}}r@{\hspace{0.4cm}}r@{\hspace{0.4cm}}r}
\hline
\hline
$L$ & $\epsilon/\tau=\frac{1}{4}$ & $\epsilon/\tau=\frac{1}{2}$ &$\epsilon/\tau=1$ & 
$\epsilon/\tau=2$& $\epsilon/\tau=4$     \\
\hline
4   &   50000  & 50000 &50000 & 50000 & 50000 \\
6   &   20000  & 20000 & 20000 &20000 & 20000 \\
8   &   15000 &13467 & 13467&  6000 &  6000 \\
10  &   10000 & 7440 & 6000  & 4918 & 4000 \\
12  &    5670  &   &  4202 & & \\
\hline
\hline
\end{tabular}
\end{center}
\caption{
Number of independent realizations of the disorder (samples) used in 
the computations.}
\label{tab_samples}
\end{table}

\subsection{Spin and link overlap}
\subsubsection{Qualitative analysis}

\begin{figure}
\begin{center}
\myscalebox{\includegraphics{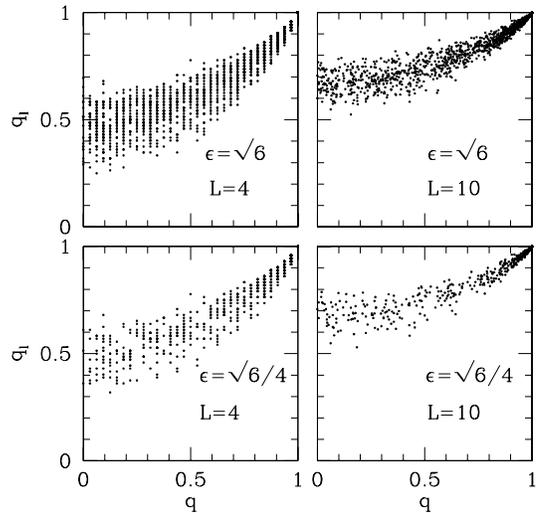}}
\caption{Scatter plots in the $(q,\ql)$ plane  for different 
values of the size  $L$ and perturbation strength $\epsilon$.
}
\label{scatter}
\end{center}
\end{figure}

We start with a qualitative analysis of the results.
In Fig.~\ref{scatter}, we show scatter plots in the 
$(q,\ql)$ plane for $L=4,10$ and 
$\epsilon/\tau={1\over 4}, 1$, where each point represents one of 2000 randomly
generated samples. Clearly, the link- and spin-overlap 
are strongly correlated.  We note that, as $\epsilon$ decreases,
there are less points with small $q$, and that as $L$ increases
the data shift towards larger values of $\ql$. Similar plots\cite{phdthesis} for periodic bc
show that $\ql$ is significantly lower than for free bc.
While $q$ has a large
variance (the points are distributed along the whole $q$ axis),
the link-overlap 
has a much smaller variance, which decreases as $L$ increases,
suggesting that in the thermodynamic limit $\ql$  either tends 
to one or becomes a well defined function of $q$.

\begin{figure}
\begin{center}
\myscalebox{\includegraphics{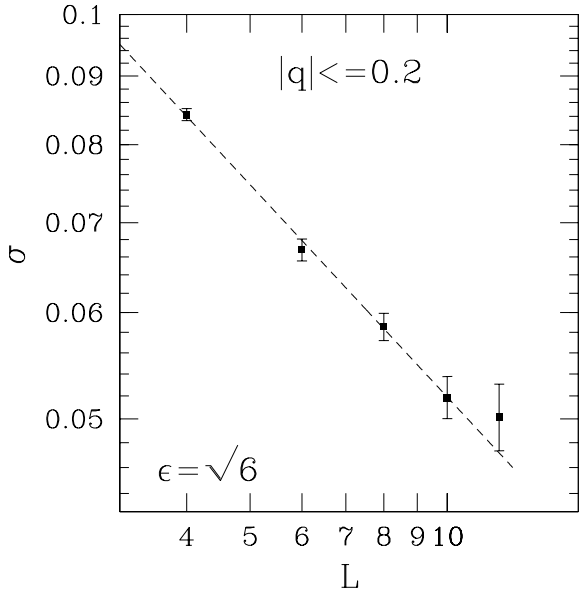}}
\caption{A plot of the standard deviation of the link-overlap
$\sigma = (\langle \ql^2\rangle_{c} - \langle \ql {\rangle}_c^2)^{1/2}$,
where the
average is restricted to samples such that $q \le 0.2$.
The line represents a power law fit with exponent 
$\delta = 0.52$.}
\label{variance}
\end{center}
\end{figure}

To quantify this, in Fig.~\ref{variance} we show
the standard deviation of $\ql$ 
\begin{equation}
\sigma = \sqrt{\langle \ql^2\rangle_{c} - \langle \ql {\rangle}_c^2 } ,
\label{sigma}
\end{equation}
restricted over samples with $q \le q_{max}$,
as a function of the system size for $\epsilon/\tau=1$.
We take $q_{\max}=0.2$, since we 
are interested in the region of  small $q$, which corresponds 
to large-scale excitations.
A power law $\sigma = a L^{-\delta}$ fits well the data with
$\delta=0.52 \pm 0.03$  ($\chi^2=1.80$, the best fit
is shown in Fig.~\ref{variance}).  Here and in the following,
unless stated otherwise, the error bars on the fit parameters
are purely statistical in relation to the fitting form considered\cite{errors}.
Restricting the average in Eq.~(\ref{sigma}) to 
different intervals of $q$ gives results also compatible with a power law.
A vanishing $\sigma$ in the thermodynamic limit is consistent with RSB,
which predicts that $\ql$ is a (nontrivial) function of $q$.
It is also trivially consistent with the 
droplet model or the  TNT picture, where $\ql=1$ for all $q$.

We also measured how the standard deviation of $q$ varies with
$L$, finding that it varies between 0.28 and 0.32. It can be fitted
both to a constant (as expected in RSB) or to a power law with 
a small exponent around 0.1. However the error bars are very large
hence the fits are not very informative. 

\begin{figure}
\begin{center}
\myscalebox{\includegraphics{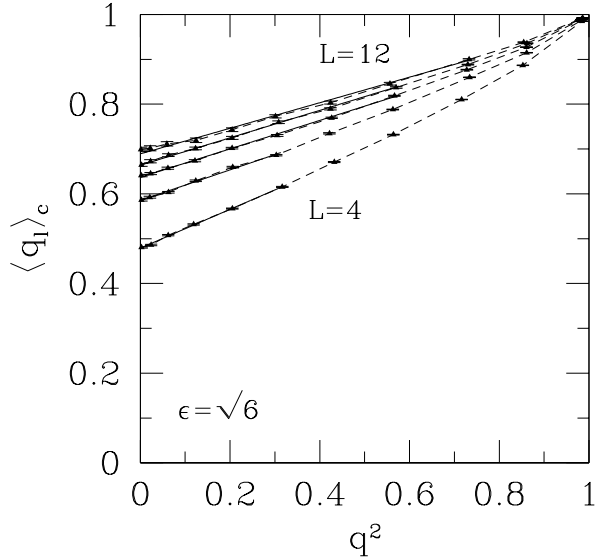}}
\caption{Average of the link-overlap restricted to intervals of $q$ of
width 0.1, as a function of $q$ for different sizes $L$ (from bottom
to top, $L=4,6,8,10,12$). The continuous
lines represent quadratic fits including values of $q$ up to where
the lines end. The dashed lines are a guide to the eye.}
\label{ql_q_L}
\end{center}
\end{figure}

Under the RSB hypothesis, it is interesting to study the functional 
relationship between $q$ and $\ql$.
In Fig.~\ref{ql_q_L} we show the {\em average}\/ value of $\ql$,
restricted to intervals $q\in [q_{\min},q_{\max}]$,
as a function of the mean value
of $q$ in each interval\cite{interval}
for $\epsilon/\tau=1$.
 For fixed $L$, a quadratic form  
$\ql=\alpha(L) + \beta(L) \,q^2$,
motivated by the infinite-range Sherrington-Kirkpatrick 
model where $\ql=q^2$,
fits well the data for $q$ less than some cut-off value which
increases with $L$ (see Fig.~\ref{ql_q_L}). The quadratic fit works 
well also for other values of $\epsilon$, and $\alpha(L)$ and $\beta(L)$ 
show a modest variation with $\epsilon$.  We also tried global fits
including data for all values of $q$ and $L$, obtaining similar results.

Extrapolating $\alpha(L)$ and $\beta(L)$ to
$L\to \infty$ with fits of the form 
$\alpha(L)=\alpha + b/L^c$, $\beta(L)=\beta + b'/L^{c'}$,
we obtain
\begin{equation}
\ql = (0.77 \pm 0.02) + (0.27 \pm 0.03) \, q^2 \, ,
\label{aval}
\end{equation}
where again the errors are purely statistical for the functional form
considered.
This nontrivial relation between $q$ and $\ql$ in the thermodynamic limit
is consistent with RSB, while in the droplet or TNT pictures the data
in Fig.\ref{ql_q_L} would shift to $\ql \equiv 1$ in this limit. 

The power law form $1- \alpha(L)=b/L^c$, $\beta(L)=b'/L^{c'}$, which 
implies $\ql=1$ in the large volume limit, fits 
poorly the data if we include all sizes.
However, if we exclude the $L=4$ data, the quality of the fit becomes
as good as that of the RSB fit just discussed. Hence, allowing for
small corrections to scaling, the droplet or TNT scenario are 
consistent with the data.

 This already shows that care must be taken to properly consider
corrections to scaling when comparing the merits of the fits  to 
various pictures.
In the following, we will investigate in detail the validity of
the various pictures by considering several observables and
explicitly discussing the corrections to scaling for each picture.

\subsubsection{Determination of $d - d_s$}

 We start with the determination of $d-d_s$ from various 
observables. We will show that for all observables, a wide range of 
values of $d-d_s$ fits well the data when allowing corrections to 
scaling, but that for all observables the smallest corrections are
attained for a value of $d-d_s$ around 0.44 as in PY.

\begin{figure}
\begin{center}
\myscalebox{\includegraphics{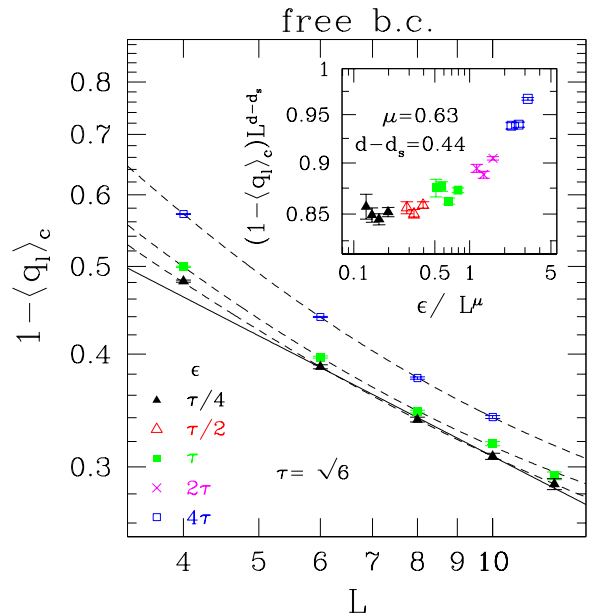}}
\caption{Logarithmic plot of the average $\la 1-\ql \ra_c$, restricted
to samples with $|q|\le 0.4$, as a function of the system size $L$. Only 
three values of $\epsilon$ are displayed for clarity.
The lower continuous line is the best fit with a power-law, 
Form III in Eq. (\ref{form1}) for
$\epsilon/\tau={1\over 4}$, where $\tau=\sqrt{6}$,
and the $L=4$ data have been excluded from the fit. 
The dashed lines are the best fits with Form I in Eq.~(\ref{form1}). 
The inset shows a scaling plot of the data in the main figure, excluding
the $L=4$ data, according to Eq.~(\ref{eqscalqlim}).
Here and in the following figures, note that the data 
for various $\epsilon$ are correlated, since the samples used for large $\epsilon$ 
were also used for small $\epsilon$.}
\label{qllim}
\end{center}
\end{figure}

The main part of Fig.~\ref{qllim} plots $\la 1- \ql \rangle_c$
as a function of $L$, for various values of $\epsilon$ and
$ q_{\max}=0.4$ ($q_{\max}=0.2$ gives essentially the same results).
First, note that, independently of what picture 
holds in the $L\to\infty$ limit, the data deviate 
significantly from asymptotic scaling, see Eq. (\ref{1mqlp}), in which the various $\epsilon$ values
should collapse on a single curve.
Second, the data have a noticeable positive (upward) curvature
for all values of $\epsilon$.
In Section~\ref{subsec:periodic} we will show that in the case of 
periodic bc the data  have a much smaller
dependence on $\epsilon$ and a much smaller curvature 
(see Fig.~\ref{qlperiodicfree}).

 In order to determine how the various pictures fit 
the data of Fig.~\ref{qllim},
 we start by considering, following Ref.\onlinecite{mapabulk}, 
 the  following three functional forms:
\begin{eqnarray}
\mbox{Form I:} \quad \la 1- \ql \ra_{c} & = & a + b/L^{c} \nonumber \\
\mbox{Form II:} \quad \la 1- \ql \ra_{c} & = & a + b/L + c/L^2  \label{form1}\\
\mbox{Form III:} \quad \la 1- \ql \ra_{c} & = & b/L^{c} \nonumber
\end{eqnarray}

Form I corresponds to the RSB prediction including the leading
correction to scaling, see Eq.~(\ref{rsbcorr}), with $c\equiv \omega$.
Form II is a different 
parameterization of the corrections to scaling.
Form III corresponds to the asymptotic behavior of both the TNT and 
droplet pictures
{\em without} corrections to scaling, see Eq. (\ref{1mqlp}), 
with $c\equiv d-d_s$.

The results of these fits (performed by $\chi^2$ minimization)
are reported in Table \ref{tablefits}.
From the Table we see that Forms I and II, appropriate to RSB,
fit well the data with a low $\chi^2$ and $a > 0$ 
outside the error bars.
The best fits with Form I are shown by the dashed lines in Fig.~\ref{qllim}.
The variation of $a$ between Forms I and II
provides a measure of the systematic
error associated with the unknown corrections to scaling.  
Within this error, $a$ is independent of $\epsilon$, as predicted by RSB.
Therefore, the data for $\la 1-\ql\ra_{c}$ are compatible
with RSB, and our central estimate under the RSB hypothesis is
\begin{equation}
\lim_{L\to \infty} \langle 1-\ql \rangle_c = 0.20 \pm 0.02 \quad \quad \mbox{(RSB)} \, ,
\label{a}
\end{equation}
where the error takes into account also the uncertainty in the {\em form}\/
of the corrections to scaling, assuming that the corrections considered 
in either Form I or II describe well the data in the whole range of sizes 
considered.
Marinari and Parisi \cite{mapabulk} fitted Form I (resp. II) to
their data for periodic 
boundary conditions, $L\le 14$, and $\epsilon/\tau=4$, and
obtained $a=0.24$ (resp. $a=0.30$),
from which we estimate a central value $a=0.27\pm 0.05$.
This is just in agreement with our estimate above for free bc,
suggesting that the infinite volume limit of $\la 1-\ql \ra_c$, 
if nonzero, may be 
independent of the boundary conditions, although we do not have an 
argument why this should be the case.

\begin{table}
\begin{center}
\begin{tabular}{l|r@{\hspace{0.3cm}}r@{\hspace{0.3cm}}r@{\hspace{0.3cm}}r@{\hspace{0.3cm}}r@{\hspace{0.3cm}}r}
\hline
\hline
Form & $\epsilon/\tau$ &  $\chi^2$      & $Q$ & $a\quad$        & $b\quad$ & $c\quad$  \\ 
\hline 
&0.25 & 0.014 & 0.99 & 	 0.171(2)  & 1.068(5) &	0.890(8) \\
&0.5  & 0.015  &0.90 &    0.185(2)  & 1.14(1)  &  0.96(1) \\
I&1 &    3.04 &  0.22 & 	0.201(14) & 1.26(7) &	1.04(7)\\
&2    & 0.39   &0.52 &    0.206(6)  & 1.42(3) &  1.08(3)\\
&4    & 0.27   &0.60 &    0.215(4)  & 1.73(3) &  1.14(2) \\
\hline
&0.25   & 0.037 & 0.98  &   0.182(1) & 1.29(2) &	-0.40(5) \\
&0.5  & 0.010  &0.92   &  0.189(1) &  1.23(1)  & -0.14(3) \\
II&1    & 3.01  & 0.22  & 0.198(8) &	1.16(10) &  0.16(27) \\
&2    & 0.47  &0.49   &  0.199(4) &  1.21(5) &  0.31(14) \\
&4    & 0.42  &0.52   &  0.202(3) &  1.31(4) &  0.64(11) \\
\hline 
&0.25& 1.40 & 0.49   &&	0.85(2) &  0.44(1) \\
&0.5  &1.63  & 0.20     & &0.87(3)  & 0.45(2)  \\
III&1 &  9.81 &  0.007 &	&0.88(3) &  0.44(2) \\
& 2    &6.75  & 0.009    &&0.95(4)  & 0.47(2)  \\
& 4    &11.4  & 0.0007   &&1.09(5)  & 0.51(2)	\\
\hline 
&0.25 &	0.55&	0.76&	0.808(4)&	7(8)&	3.6(8) \\
&0.5 &	0.65&	0.42&	0.811(6)&	6(7)&	3.2(8)\\
IV& 1&	5.91&	0.05&	0.828(7)&	18(37)&	4.0(1.5)\\
&2&	2.80&	0.09&	0.844(9)&	5(5)&	2.9(7)\\
&4&	3.01&	0.08&	0.87(1)	&	3.8(1.6)&	2.3(3)	\\
\hline
\hline
\end{tabular}
\end{center}
\caption{ 
Fits to $\la 1- \ql \ra_{c}$ with $q_{\min}=0$ and $q_{\max}=0.4$. The four groups
of data refer, from top to bottom, to the three fitting functions I, II, and III 
in Eq. (\ref{form1}), and Form IV in Eq. (\ref{form4}) with $d-d_s=0.44$.
For Form III,
data for $L=4$ was not included in the fit. The errors are the
standard errors of a nonlinear fitting routine\cite{errors}, and
 $Q$ is the goodness-of-fit parameter.}
\label{tablefits}
\end{table}

The power law Form, III, appropriate to the droplet model or the TNT scenario,
does not fit well the data if we include all the sizes,
but if we exclude $L=4$, it does fit well for $\epsilon/\tau < 2$,
and the fit parameters $b$ and $c$ are almost independent of $\epsilon$. 
The quality of the fit of Form III is worse than that of Forms I and II, 
but still acceptable.
 The main point we want to stress, however, is that the worse fit of 
Form III alone does {\em not} necessarily favor the RSB picture, since Form III 
does not include corrections to scaling, while Forms I and II do.
In other words, Forms I and II are rather ``forgiving''
with the RSB picture, allowing corrections of magnitude
100$\%$ - 200$\%$ of the predicted $L$-independent asymptote,
while Form III demands that the power-law scenario fits
with corrections smaller than the (very small) statistical errors.
By looking at Figure~\ref{qllim}, it is clear that the
data are closer to a power law than to an $L$-independent behaviour.

Therefore, in order to try a comparison that puts the various pictures 
on an equal footing,
we performed fits with the following more general functional form:
\begin{equation}
\mbox{Form IV:} \quad L^{d-d_s}  \la 1- \ql \ra_{c}  =  a + b/L^{c} \label{form4}
\end{equation}
where we fix $d-d_s$ and minimize the $\chi^2$ with respect to $a,b,c$,
repeating the procedure for different values of $d-d_s$.
For $d-d_s=0$, Form IV reduces to Form I, while for $d-d_s > 0$, it
corresponds to Form III plus a correction-to-scaling term, with 
correction-to-scaling exponent $\omega=c$. We find that, as we might have 
expected  from the previous discussion, Form IV fits well 
the data for  a wide range of values of $d-d_s$. For example, for 
$\epsilon/\tau={1\over 4}$, 
a value of $d-d_s$ between 0 and 0.45 gives a goodness-of-fit parameter 
$Q \ge 0.43$, which is entirely acceptable. 

This shows that, when allowing for corrections to scaling for
{\em all} pictures, the droplet or TNT pictures are as good as RSB 
as far as the statistical quality of the fits is concerned.
However, within the interval of acceptable values
of $d-d_s$, clearly the larger is $d-d_s$ the smaller are the corrections to 
asymptotic scaling. For example, for
$\epsilon/\tau={1\over 4}$ and $d-d_s=0.42$,
the correction term $b/L^{c}$ in Form IV amounts to 
only $0.1\%$ of the total for $L=12$, while for $d-d_s=0$ it amounts
to $43\%$. 
Hence a large value of $d-d_s$
may be regarded as more ``natural'' in this range
of sizes.

If we impose 
that the correction to scaling for $L\ge 8$ is less than an (arbitrary) 
limit of $3\%$,  we obtain  the estimate 
\begin{equation}
d-d_s=0.44 \pm 0.03
\label{dds}
\end{equation}
where the error is purely statistical within this assumption.
In Table II we show the fits obtained
with Form IV imposing this value. This agrees with the estimate
$d-d_s=0.42 \pm 0.02$ of PY for periodic bc (see also
Section~\ref{subsec:periodic} of
this paper), which is  reassuring since $d-d_s$ should not
depend on the boundary conditions.  

Note that for $d-d_s=0$, corresponding to Form I, corrections within $3\%$ from 
the asymptotic limit would only be attained for a size $L\simeq 200$.
We also note that, 
as we discussed in Section \ref{sec:method}, 
even in the regime where corrections  to scaling are negligible,
asymptotic scaling sets in only for $L \gg \epsilon^{1/\mu}$.
This explains why, if $d-d_s \simeq 0.44$,
the quality of the power-law fit in Table II becomes progressively worse as 
$\epsilon$ increases.  
In particular, the deviation from  asymptotic scaling
is very significant for $\epsilon/\tau=4$, and hence
from the data of $\epsilon/\tau=4$ alone one should
not necessarily conclude\cite{mapabulk} that 
an asymptotic power-law behavior is ruled out.
This is seen also in the inset of Fig.~\ref{qllim}, 
which shows that, if we exclude $L=4$,
the data are compatible with the scaling relation Eq.~(\ref{scaling}), 
where the exponent
$\mu$ is independently determined below.

PY determined $d-d_s$ from the ratio $R=\la 1-\ql\ra/\la 1- q\ra$
which has the same scaling behavior as the quantity $\la 1-\ql \ra_{c}$
used here, namely $R =   L^{-(d-d_s)} F_R(\epsilon/L^\mu)$, with
$F_R(x) \sim $ const. as  $x\to 0$.
Middleton \cite{middleton} observed that, in two dimensions, small droplets introduce 
significant corrections to scaling, and  suggested 
that this may be the case also in three dimensions, possibly invalidating 
the conclusions of PY. The quantity $\langle 1-\ql \rangle_c$ is less 
sensitive to these corrections since, by restricting to small $q$, small 
droplets should give a smaller contribution, 
because we have eliminated the part at large $q$ where one can
have {\em only}\/ small droplets.
Hence, to investigate these corrections, we fitted our data for $R$
with  Forms I-IV above (with $R$ replacing
$\la 1-\ql \ra_c$). The results we find are very similar to those for
$\la 1-\ql \ra_c$: Forms I and II fit well the data with a low
$\chi^2$, giving $a=0.27 \pm 0.03$ independent of $\epsilon$
within the error bars. A power law fits well the data if we exclude $L=4$,
with an exponent  $d-d_s=0.43 \pm 0.03$ nearly independent of $\epsilon$ 
and in agreement
with Eq.~(\ref{dds}). The residual dependence on $\epsilon$ 
is well accounted for by a scaling plot similar to the inset
in Fig.~\ref{qllim}.  Form IV also
fits well the data for a wide range of values of $d-d_s$.
 Again, a power law is more natural in the sense
that corrections to scaling are smaller, and the smallest corrections
are obtained for $d-d_s$ around 0.43 as for $\langle 1-\ql \rangle_c$.
We interpret the fact that the two quantities give the same
value of $d-d_s$ as evidence that corrections due to small droplets are 
indeed not important in three dimensions in this range of sizes. 
In Section \ref{subsec:periodic} we will show that this is also
the case for periodic bc.

To summarize this part, the data  for both $R$ and $\la 1-\ql\ra_c$ 
are compatible with a wide range of values of $d - d_s$ between zero (corresponding
to RSB) and $\simeq 0.44$, but a value at the higher end of this range
describes the data in a more natural way, in the sense that the corrections to
scaling are smaller.

\subsubsection{Determination of  $\theta'$}

 Next, we turn to the exponent  $\mu$ defined
in Eq.~(\ref{mudef}), from which we will extract the exponent $\theta'$ 
which is the other exponent, with $d-d_s$, characterizing the spin glass phase.
To this end we consider the  ratio
\begin{equation}
B = { \langle 1-\ql \rangle^2 \over \langle (1-\ql)^2 \rangle}
\label{ratioB}
\end{equation}
which follows the scaling law
\begin{equation}
B =   F_{B}(\epsilon/L^\mu) .
\label{scalingB}
\end{equation}
The factor $L^{d-d_s}$ 
does not appear here since it 
cancels between numerator and denumerator of Eq.~(\ref{ratioB}), thus
allowing us to determine $\mu$ independently of $d-d_s$.
Following the arguments in Section~\ref{sec:method}, we expect
${F_{B}}(x) \sim x$ for small $x$ since both $L^{d-d_s} \langle 1-\ql \rangle$ and
$L^{2 (d-d_s)} \langle (1-\ql)^2 \rangle$ vary as $\epsilon/L^\mu$ for $\epsilon/L^\mu \to 0$, 
hence the asymptotic scaling of $B$ is $B\sim \epsilon/L^\mu$.

\begin{figure}
\begin{center}
\myscalebox{\includegraphics{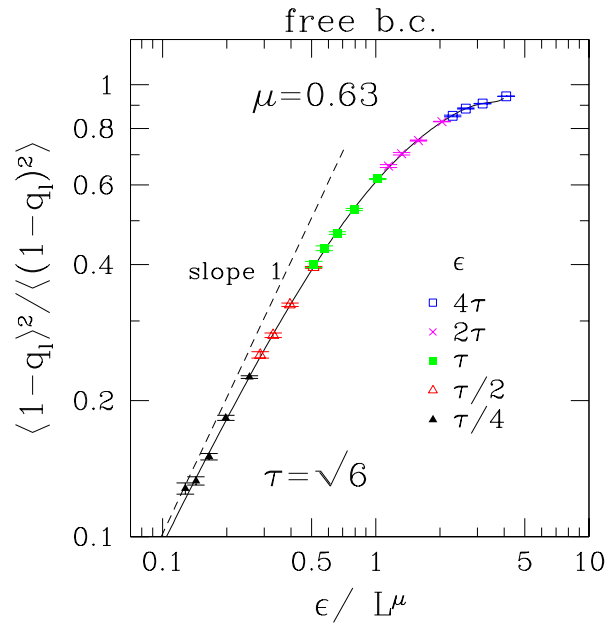}}
\caption{Scaling plot of the ratio $B = \langle 1-\ql \rangle^2 / \langle (1-\ql)^2 \rangle$
 according to Eq.~(\ref{scalingB}).
The continuous line is a polynomial fit of order $n=6$, which gives
$\chi^2/$d.o.f = 1.09, and a goodness-of-fit parameter $Q=0.35$. 
The dashed line
is the linear term of the polynomial fit, corresponding to the
asymptotic scaling for $L\to \infty$. }
\label{scalingratioql}
\end{center}
\end{figure}

 To determine $\mu$, we fit the scaling law Eq. (\ref{scalingB}) 
to our data assuming a polynomial form of order $n$ for $F_B(x)$, 
 namely $F_B(x)=\sum_{i=0,n} c_i\, x^i$, with $c_0$ set to zero 
in order to satisfy the asymptotic scaling  $F_B(x)\sim x$ as $x\to 0$.
We repeat the fit in an interval of values for $\mu$, and 
determine the value of $\mu$ which gives the best fit, varying 
$n$ until the $\chi^2$ of the best fit 
becomes approximately constant. In this way we obtain 
\begin{equation}
\mu=0.63 \pm 0.03  ,
\label{estimatemu}
\end{equation}
 where the error is purely statistical, under the assumption that the
corrections to scaling are smaller than the statistical errors of the data.
 As shown in Fig.~\ref{scalingratioql}, scaling is quite satisfactory, 
 with all the data collapsing on one curve, 
although the data for different $\epsilon$ overlap only slightly. 
The best fit for $n=6$ is displayed by the continuous line.
We emphasize that this scaling plot is obtained with only {\em one}\/
adjustable parameter, $\mu$.
 Note that in the {\em asymptotic}\/ scaling regime the data 
should follow a straight line (power-law), while the data in the figure
show a pronounced curvature.
Significant
corrections to asymptotic scaling must be expected for large $\epsilon$, since
$B$ must satisfy the inequality $B \leq 1$. 
The dashed line in Fig.~\ref{scalingratioql}
represents the linear term of $F_B(x)$, corresponding to asymptotic scaling,
and the deviation from it
gives a measure of the corrections to asymptotic scaling.
The $\epsilon/\tau=
{1\over 4}$ data are quite close to asymptotic scaling, while the data 
for large $\epsilon$ deviate significantly from it.  Another manifestation
of these corrections is that, if we fit the data with a power-law 
$B=b/L^{\tilde{\mu}}$, the effective exponent $\tilde{\mu}$
 varies strongly with
$\epsilon$, converging towards 0.63 for $\epsilon\to 0$. 

In RSB, $B \sim \epsilon$ as $L\to\infty$ since $\mu=0$.
To test the RSB prediction, we performed fits of $B$ using
  Form I and Form II in Eq.~(\ref{form1}),
(where $\la 1-\ql \ra_c$ is replaced by  $B$).
Form I gives unphysical
(negative) values of $a$, while Form II gives an acceptable fit with a positive
$a$ roughly proportional to $\epsilon$. Therefore, the data for $B$ cannot 
rule out RSB. 
Note that, if RSB holds asymptotically, the data in  Fig.\ref{scalingratioql}
would deviate from the scaling curve for larger $L$, saturating to a constant
value for small values of $\epsilon/L^\mu$. The good data collapse we observed,
therefore, would be entirely accidental.

We believe that the observed data collapse 
is a good indication towards the validity of a scaling scenario with 
large $\mu$. Certainly this scenario is more natural since it fits the data with
(almost) no corrections to (simple) scaling, while the corrections for RSB 
are very large as apparent from Fig.\ref{scalingratioql}. 
A similar conclusion was reached in the determination of
$d-d_s$.

 As a further  test,
we can obtain a second estimate of $\mu$ from
the quantity $\la 1-\ql \ra$ ($q$  unrestricted), whose
scaling and asymptotic scaling are given 
in Eqs.~(\ref{scaling}) and (\ref{1mql}).
 We find results similar to those for $B$:
A power law fit $\la 1-\ql \ra = b/L^{\tilde{\mu_l}}$ (Form III) 
gives acceptable fits for $\epsilon/\tau \le {1\over 2}$ and for all values 
of $\epsilon$ if we exclude $L=4$.  As for $B$, the 
effective exponent $\tilde{\mu_l}$ changes significantly 
with $\epsilon$, due to corrections to  asymptotic scaling, and by 
extrapolating it to $\epsilon = 0$ we obtain $\mu_l = 1.15 \pm 0.08$.
This gives
$\mu=\mu_l - (d-d_s)=0.60 \pm 0.09$
which agrees with the estimate $\mu = 0.63 \pm 0.03$ 
obtained from $B$. 
We also verified that, as for $B$, the data collapse
reasonably well on one curve for $\mu = 0.64 \pm 0.05$
according to Eq.~(\ref{scaling}),
although the quality of the scaling is somewhat worse than that
of Fig.\ref{scalingratioql}. 
To check the RSB prediction, we fitted the data to Forms I and II
 (where now $\la 1-\ql \ra_c$ is replaced by  $\la 1-\ql \ra$),
finding that they both fit well the data, with $a$ roughly proportional
to $\epsilon$ as expected in RSB, although, for small $\epsilon$, $a$ is
also compatible with zero.  Therefore, as for $B$, 
the data are also consistent with RSB, but
this scenario requires large corrections to scaling, 
while the hypothesis $\mu = 0.63$
fits the data with almost no corrections to (simple) scaling.

In the analysis so far, we have determined the exponents $\mu$ and $d-d_s$
using just the link-overlap $\ql$. By contrast, PY determined $\mu$ 
(for periodic bc) from the scaling of the spin-overlap $q$.
An advantage of $\ql$ is that its variance is much lower, as shown 
in Fig.~\ref{scatter}. In any event, we have verified that the scaling relation
 Eq.~(\ref{scalingq}) fits well the data for $q$,
giving $\mu = 0.65 \pm 0.02$, in agreement with 
the estimates from $B$ and $\la 1-\ql \ra$. 

Summarizing this part, we find that the data for  {\em all} the quantities
considered, namely $B$, $\la 1-\ql \ra$, and
$\la 1-q \ra$, are consistent with the RSB
prediction that $\mu = 0$ asymptotically, but large corrections to scaling are
required in the fit,  similarly to what we observed in the
determination of $d-d_s$. The data are also fitted
very well by a scaling scenario with $\mu \simeq 0.63$, with almost
negligible corrections to scaling (but with sizeable corrections to asymptotic
scaling, which instead were small for the observables 
considered for $d-d_s$). 
Under the ``natural'' assumption of small corrections to scaling, 
from the estimates of $\mu$ and $d-d_s$ 
in Eqs.~(\ref{estimatemu}) and (\ref{dds}), we obtain
\begin{equation}
\theta'= \mu - (d-d_s) = 0.19 \pm 0.06 ,
\end{equation}
 where, again, the error is purely statistical subject to the condition 
of having small (less than 3\%) corrections to scaling.
This result agrees with the droplet theory which predicts that
$\theta' = \theta >0$,
and is compatible with the value of $\theta \simeq 0.2$ 
characterizing the  energy of domain walls
induced by a change in boundary conditions\cite{theta-3d}.
By contrast, for periodic bc  and under the same assumption of small
corrections to scaling, $\theta'$ is compatible with 
zero (see PY and Section~\ref{subsec:periodic}).
In Section \ref{subsec:discussion} we will analyze
the origin of this discrepancy, and show that by allowing small (of order 10$\%$)
corrections
to scaling the free bc data can be reconciled with $\theta'\simeq 0$.

\subsection{Box overlaps}

\begin{figure}
\begin{center}
\myscalebox{\includegraphics{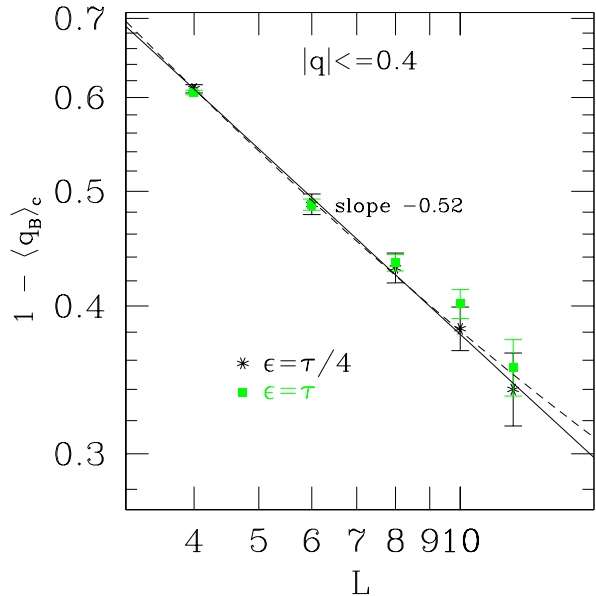}}
\caption{Logarithmic plot of the average  box-overlap, restricted to 
samples such that $q\le 0.4$.
We show the data for just two values of $\epsilon$ for clarity. The data 
for other
values of $\epsilon$ are superimposed. The lower continuous line is a
power-law fit for $\epsilon/\tau=4$. The dashed line is the fit with
Form II in Eq. (\ref{form1}), with $q_B$ replacing $\ql$. The slope gives
the exponent $d-d_s$.}
\label{boxlim}
\end{center}
\end{figure}

 So far we have analyzed the link- and spin-overlap which are computed
on the whole system (bulk). 
We now turn to a different observable, the {\em box-overlap}\/ defined as
\begin{equation}
q_B = {1\over L_B^d} \sum_i S^{(0)}_i  \tilde{S}^{(0)}_i 
\label{def_qb}
\end{equation}
where the sum runs over the sites contained in a central cubic
box of {\em fixed}\/ size $L_B=2$. In the following we will only consider 
the absolute value $|q_B|$, which we still call $q_B$ for simplicity.
When a large-scale cluster of spins is flipped, for large $L$ the probability 
that its surface goes across the central box is proportional
to  the ratio of its surface area, $\sim L^{d_s}$,
to the volume, $L^d$. Therefore $1-q_B \sim L^{-(d-d_s)}$ 
from which we obtain the scaling laws
\begin{eqnarray}
\la 1 - q_B \rangle &=&  L^{-(d-d_s)} F_{q_B}(\epsilon/L^\mu) 
\label{eqscalqb} \\
\la 1 - q_B \rangle_c &=&  L^{-(d-d_s)} F^c_{q_B}(\epsilon/L^\mu) 
\label{eqscalqblim}
\end{eqnarray}
where, as for the corresponding scaling functions for $\ql$,
${F_{q_B}}(x)\sim x$ and ${F^c_{q_B}}(x)\sim $ const. for small $x$.
Hence the asymptotic scaling for $L\to \infty$ is
\begin{eqnarray}
\la 1 -  q_B \rangle &\sim& \epsilon L^{-\mu_l}
\label{asymptqb} \\
\la 1 -  q_B \rangle_c &\sim&  L^{-(d-d_s)} \, .
\label{asymptqblim}
\end{eqnarray}
In RSB, this reduces to $\la 1 -  q_B \rangle \sim \epsilon$ and 
$\la 1 -  q_B \rangle_c \sim$ const.
An advantage of $q_B$ over $\ql$ is that the former, being measured
away from the boundaries, should have smaller corrections to scaling
and be less sensitive to boundary conditions. 
Indeed, Monte Carlo 
simulations\cite{ricci,py_unpub} show that $q_B$ has rather
small corrections to scaling. This is likely to be
particularly important for the free boundary conditions used here.

Fig.~\ref{boxlim} shows the {\em restricted}\/ average
$\la 1 -  q_B \rangle_c$, with $q_{\max}=0.4$,
as a function
of $L$ for two values of $\epsilon$. The data are clearly decreasing
with $L$, are essentially independent of $\epsilon$, as expected
from Eq.~(\ref{asymptqblim}), and are  close to a straight line on the
logarithmic plot. This indicates that the power law fit, Form III, appropriate to
the droplet and TNT scenarios, should work well and indeed it does, even for
the largest
value of $\epsilon$
(we note however that the statistical errors are larger than 
for the link-overlap,  hence the fits are less sensitive to 
corrections to scaling). The exponent is almost independent of $\epsilon$,
varying between 0.48 and 0.52, and from this we obtain the estimate
\begin{equation}
d-d_s = 0.48 \pm 0.03
\label{dds_box}
\end{equation}
which is in agreement with the estimates $d-d_s = 0.44 \pm 0.03$ from
$\la 1-\ql \ra_c$ and  $d-d_s = 0.43 \pm 0.03$ from $R$.

Forms I and II (with $q_B$ replacing $\ql$)
also fit well the data, with $a$ 
between 0.14 and 0.36 (with no
discernible trend with $\epsilon$). Hence the data are also compatible
with RSB, and under the RSB hypothesis, we estimate
\begin{equation}
\lim_{L\to \infty} \langle 1-q_B \rangle_c = 0.25 \pm 0.10 \quad \quad \mbox{(RSB)} \,.
\label{limqB}
\end{equation}
 As usual, we note that the RSB scenario requires rather large corrections
to scaling, while the power law fits the data with no corrections.

\begin{figure}
\begin{center}
\myscalebox{\includegraphics{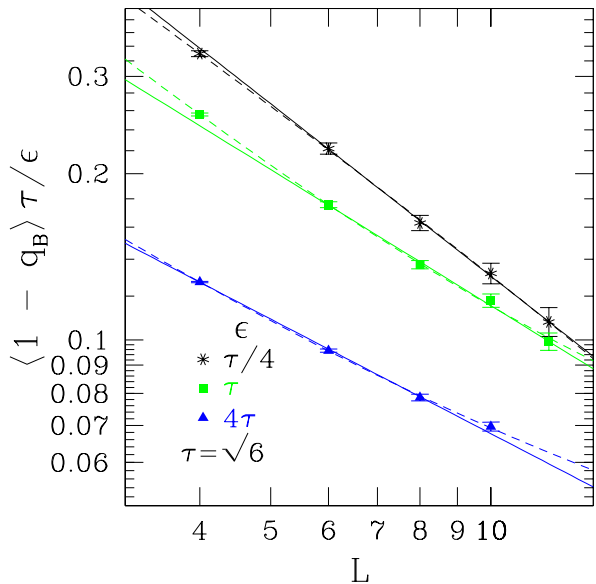}}
\caption{Logarithmic plot of the average box-overlap, multiplied by
$\tau/\epsilon$  in order to highlight the deviation from
the asymptotic behavior of Eq.~(\protect\ref{asymptqb}) in which
the data for various $\epsilon$ should collapse on a single curve. 
The continuous lines represent fits with the power-law Form III
excluding $L=4$. The dashed lines represent fits with Form I in 
Eq. (\ref{form1}).}
\label{box}
\end{center}
\end{figure}

\begin{figure}
\begin{center}
\myscalebox{\includegraphics{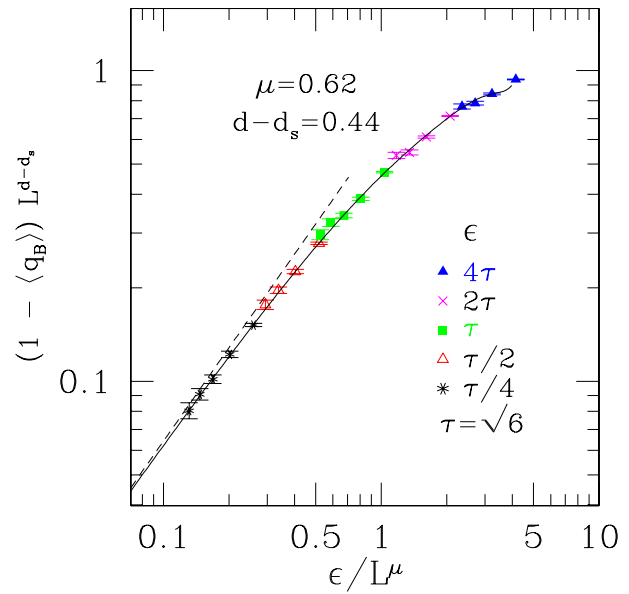}}
\caption{Scaling plot of the box-overlap according to Eq.~(\ref{eqscalqb}).
The continuous line is a polynomial fit of order $n=6$, which gives 
$\chi^2/$d.o.f = 0.63,  and a goodness-of-fit parameter $Q=0.85$. The dashed line
is the linear term of the polynomial fit, corresponding to the
asymptotic behavior for $L\to \infty$.}
\label{scalingbox}
\end{center}
\end{figure}

Fig.~\ref{box} shows the  unrestricted average $\la 1 -  q_B \rangle$
multiplied by $\tau/\epsilon$, which asymptotically should be independent of
$\epsilon$. The data show a small curvature and
a significant $\epsilon$ dependence, 
indicating that for this quantity
we are not yet in the asymptotic scaling regime (similarly to what we 
observed for the quantity $B$).
The data are fitted
well by a power law, with an exponent that changes with $\epsilon$ and
tends towards $\mu \simeq 0.63$ for $\epsilon \to 0$.
Fits using Forms I and II
give $a$ compatible with zero.  We also determined $\mu$ from 
the scaling relation Eq. (\ref{eqscalqb}), by fixing $d-d_s=0.44$ and
using the same fitting procedure as for $B$ (which assumes no
corrections to scaling), finding 
\begin{equation}
\mu=0.62 \pm 0.04
\end{equation}
which agrees with the various estimates of $\mu$ obtained 
from $B$, $\la 1-\ql\ra$, and $\la 1-q\ra$. Fig.~\ref{scalingbox}
shows the corresponding scaling plot,  in which 
the data collapse is fairly good.

To conclude this subsection, the data for box overlaps can be fitted with
smaller corrections to scaling than the data for the bulk link- and spin-overlap.
A fit to the generic scaling picture,
with no corrections to scaling, gives
results for the exponents $d-d_s$ and $\mu$ in
 agreement with those from the bulk quantities analyzed in the previous 
subsections. 
However, as with the bulk observables, 
assuming large corrections to scaling, the data can
also be fitted to the RSB picture.

\subsection{Comparison with periodic boundary conditions}
\label{subsec:periodic}

In order to assess the effect of different 
boundary conditions, we have repeated part of 
the analysis above (with the exclusion of box-overlaps)
for the data of PY (Ref.\onlinecite{py-bulk}) for periodic 
boundary conditions and $L\le 8$. The ground states were obtained 
using a hybrid genetic
algorithm as described in PY. This does not
guarantee to find the true ground state, but the systematic
errors due to occasionally missing it are smaller than the
statistical errors\cite{py-bulk}.

\begin{figure}
\begin{center}
\myscalebox{\includegraphics{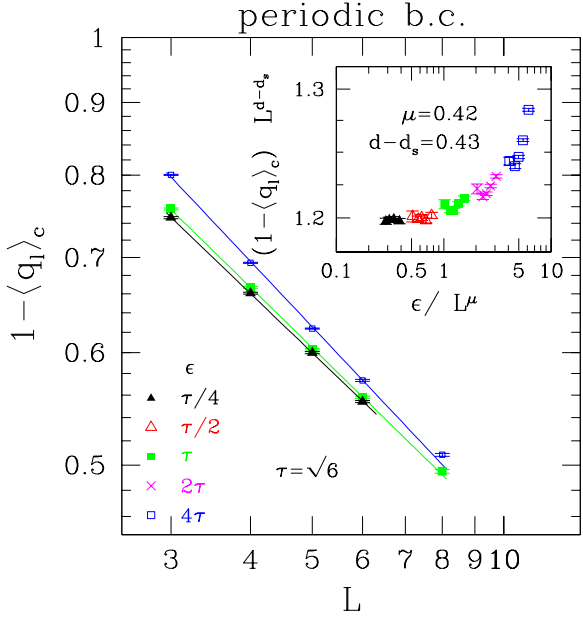}}
\caption{Same as Fig.~\ref{qllim} but for periodic boundary conditions,
using the data of PY (Ref.\protect\onlinecite{py-bulk}).}
\label{qllimper}
\end{center}
\end{figure}

\begin{figure}
\begin{center}
\myscalebox{\includegraphics{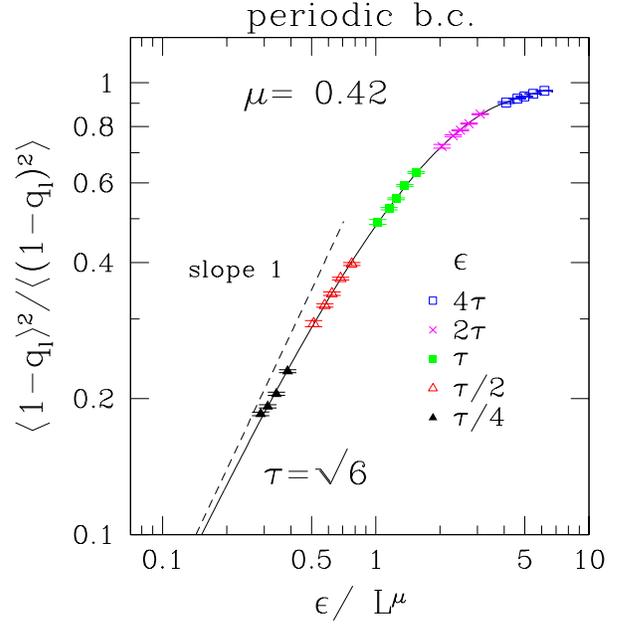}}
\caption{Same as Fig.~\ref{scalingratioql} but for periodic boundary conditions,
using the data of PY (Ref.\protect\onlinecite{py-bulk}).}
\label{ratioper}
\end{center}
\end{figure}

Figs.~\ref{qllimper} and \ref{ratioper}
show the equivalent for periodic bc of Figs.~\ref{qllim} and \ref{scalingratioql}  
for free bc.
The data in Fig.~\ref{qllimper} shows much less curvature  
and also a smaller dependence on $\epsilon$ than
the corresponding data for free bc in Fig.~\ref{qllim}, indicating that
corrections to scaling are smaller for periodic bc.
Table \ref{tablefitsperiodic} reports the best fits using the three functional 
forms of Eq. (\ref{form1}).
Form I fits well the data, but $a$ varies significantly
with $\epsilon$, and for small $\epsilon$ it is compatible with zero.
Form II also fits well, with $a$ independent of $\epsilon$
within the statistical errors. From this fit we estimate
\begin{equation}
\lim_{L\to \infty} \langle 1-\ql \rangle_c = 0.28 \pm 0.03 \quad \quad \mbox{(RSB)} 
\label{a_per}
\end{equation}
 (see the comment after Eq.(\ref{a}) as to the meaning of the error bar)
which agrees with the estimate 0.24 of Marinari and Parisi\cite{mapabulk},
and is just consistent with our estimate $0.20\pm 0.02$ for free bc.

The power-law fit with no corrections, Form III,
fits well the data for the two smallest values of
$\epsilon$ and, if we exclude $L=3$, for all but the largest value of
$\epsilon$.
The exponent $c \equiv d - d_s$ is nearly independent of $\epsilon$ and gives
\begin{equation}
d-d_s=0.43 \pm 0.02  \quad {\mbox{(periodic bc)}} \,.
\label{dds_per}
\end{equation}
This result agrees with the estimate $d-d_s=0.42\pm 0.02$ of PY obtained
from the ratio $R$ defined above, confirming that corrections due to small
droplets should not be important in three dimensions, and with
our estimate $d-d_s=0.44 \pm 0.03$ for free bc, indicating that $d-d_s$ does
not depend on boundary conditions.

We also performed fits with Form IV which includes corrections to scaling.
 As for free bc,  a wide range of values of $d-d_s$ from
zero to around 0.44 give a good fit, with the largest values giving the
smallest corrections to scaling.
The results  of the fit
for $d-d_s=0.43$ are shown in Table \ref{tablefitsperiodic}.
For the two smaller values of $\epsilon$, the fits are difficult because
corrections to scaling are very small,  hence they are not shown.

\begin{table}
\begin{center}
\begin{tabular}{l|r@{\hspace{0.3cm}}r@{\hspace{0.3cm}}r@{\hspace{0.3cm}}r@{\hspace{0.3cm}}r@{\hspace{0.3cm}}r@{\hspace{0.3cm}}r}
\hline
\hline
Form & $\epsilon/\tau$ &  $\chi^2$      & $Q$ & $a$        & $b$ & $c$  \\ 
\hline 
&0.25 & 0.003 & 0.99  &  -0.076(7) & 1.256(6) &  0.384(4)\\
&0.5  & 0.92  & 0.62  &   0.05(4)  & 1.16(2) &  0.47(3)\\
I&1    & 1.58  & 0.45  &   0.10(3)  & 1.16(2) &  0.52(3)\\
&2    & 2.20  & 0.33  &   0.12(3)  & 1.18(2) &  0.54(4)\\
&4    & 1.58  & 0.45  &   0.20(2)  & 1.270(4) &  0.68(2)\\
\hline 
&0.25 & 0.33 &  0.56  &   0.279(7) & 1.90(5)  &-1.5(1)\\
&0.5  & 5.22 &  0.073 &   0.28(1)  & 1.90(9)  &-1.5(2)\\
II&1    & 0.69 &  0.71  &   0.280(4) & 1.89(3)  &-1.40(6)\\
&2    & 0.04 &  0.98  &   0.283(1) & 1.90(1)  &-1.33(2)\\
&4    & 0.36 &  0.83  &   0.291(2) & 1.86(2)  &-1.01(4)\\
\hline 
&0.25  &  0.02	&0.887	&&1.204(2)&	 0.433(1)\\
&0.5 &	0.35	&0.838	&&1.193(3)&      0.427(2)\\
III&1	&5.14	&0.076	&&1.21(2)&	 0.434(6)\\
&2	&7.82	&0.020	&&1.24(2)&	 0.440(8)\\
&4  &25.7 &2 $10^{-6}$&&	  1.31(2)	&0.46(1)\\
\hline 
&1	&3.59	&0.16	&1.205(4)	&0.3(1.5)&3(4)\\
IV&2	&4.69	&0.09	&1.214(7)	&0.2(5)	&2(2)\\
&4	&8.38	&0.01	&1.231(8)	&0.7(5)&	2.3(7)\\
\hline
\hline
\end{tabular}
\end{center}
\caption{ 
Fits to $\la 1- \ql \ra_{c}$ with $q_{\min}=0$ and $q_{\max}=0.4$ for
periodic boundary conditions. The three groups
of data refer, from top to bottom, 
to the three fitting functions I, II, and III 
in Eq. (\ref{form1}), respectively, and Form IV in Eq. (\ref{form4}) with $d-d_s=0.43$.}
\label{tablefitsperiodic}
\end{table}

We determined the exponent $\mu$
from the ratio $B$ using Eq. (\ref{scalingB}) and
the fitting procedure described for free bc, obtaining 
\begin{equation}
\mu=0.42 \pm 0.03 \quad  {\mbox{(periodic bc)}}
\label{estimatemu_per}
\end{equation}
 where, as for the estimate of $d-d_s$ above, the errors 
are purely statistical with the assumption that
corrections to scaling are smaller than the statistical errors of
the data. Scaling is rather satisfactory 
as shown in Fig.~\ref{ratioper}. This value 
agrees with the estimate $\mu=0.44\pm 0.02$ of PY from 
the scaling of the spin-overlap 
but incompatible, within the statistical 
error bars, with the result $\mu=0.63 \pm 0.03$ for free bc.
 We will return in Section ~\ref{subsec:discussion} 
on the origin of the discrepancy between free and periodic bc.
The inset of Fig.~\ref{qllimper} shows that, with these values of
$\mu$ and $d-d_s$, the scaling form for $\la 1-\ql \ra_c$, 
Eq.~(\ref{eqscalqlim}), is also well satisfied.
Finally, we verified that, if $d-d_s=0.43$, the unrestricted 
average $\la 1-\ql \ra$ satisfies  scaling,
giving $\mu=0.45 \pm 0.02$ in agreement with  the estimate from $B$.

Combining Eqs.~(\ref{dds_per})
and (\ref{estimatemu_per}),  we obtain the estimate
of $\theta'$ for periodic bc:

\begin{equation}
\theta' = \mu - (d-d_s) = -0.01 \pm 0.03 \quad {\mbox{(periodic bc)}} \, .
\end{equation}

This is compatible with zero and, within the error bars, 
incompatible with the value $\theta'=\theta \simeq 0.2$, where $\theta$
characterizes
the energy of domain walls induced by boundary condition changes. 
A scenario in which $\theta'=0$ and $d-d_s > 0$ is consistent 
with the TNT picture.  Finally, we note that, 
although our analysis of the PY data uses 
different quantities to extract exponents, our results agree
with those given by PY.

\subsection{Discussion and summary of the results}
\label{subsec:discussion}

 In the previous sections, we have seen that for both free and
periodic bc, the analysis of 
all the different observables considered gives consistent results for the 
exponents 
$d-d_s$ and $\theta'$ under the assumption of minimal corrections to
scaling. However, 
while the results for $d-d_s$ for free and periodic bc agree with each other,
the results for $\theta'$ apparently do not, having found 
$\theta'=-0.01 \pm 0.03$ for periodic bc and $\theta'=0.19\pm 0.06$
for free bc.
Since $\theta'$,  like $d-d_s$, should not depend on the 
type of boundary conditions, the discrepancy must be due to 
different corrections to scaling for the two  boundary conditions.

Therefore, it is important to analyze further the corrections
to scaling. First, we recall that the scaling plots for the quantity $B$
in Figs.~\ref{scalingratioql} (free bc)  and \ref{ratioper} (periodic bc), 
from which we have
determined $\mu$ (and hence $\theta'$), are obtained by imposing 
that the {\em whole} data set (namely all values of $\epsilon$ and $L$)
satisfies scaling with corrections to scaling
smaller than the statistical errors, 
which are less than $1\%$. Clearly, this is a very stringent requirement.
If we relax this requirement, allowing some corrections to (simple)
scaling, we can accommodate a larger range of values for $\mu$.

\begin{figure}
\begin{center}
\myscalebox{\includegraphics{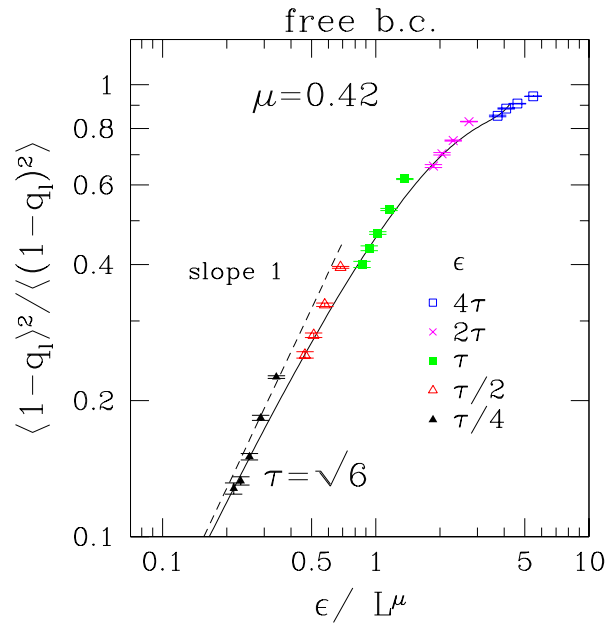}}
\caption{Scaling plot of the ratio $B = \langle 1-\ql \rangle^2 / 
\langle (1-\ql)^2 \rangle$
 according to Eq.~(\ref{scalingB}).
The continuous line is a polynomial fit of order $n=5$, excluding
the data with $L=4$ and $6$,
which gives $\chi^2/$d.o.f $= 1.26$.
The dashed line
is the linear term of the polynomial fit, corresponding to the
asymptotic scaling for $L\to \infty$. }
\label{scalingratioql42}
\end{center}
\end{figure}

This is shown in Fig.~\ref{scalingratioql42}, which gives
a scaling plot for free bc,
analogous to Fig.~\ref{scalingratioql} but assuming the value 
$\mu=0.42$ determined from {\em periodic}\/ bc. The polynomial 
fitting curve was obtained by excluding from the fit the data points 
for $L=4$ and 6. One can see that for larger $L$ the data collapse
reasonably well on the curve. The deviation of the $L=4,6$ data from 
the curve, less than $10\%$, is a measure of the corrections to scaling.
Therefore, we see that corrections to scaling of less than $10\%$ for
the two smallest sizes are sufficient to remove the discrepancy
in the value of $\theta'$ between free and periodic bc.
We verified that the also the other quantities 
considered, namely $\la 1-\ql \ra_c$, $R$, $\la 1-\ql \ra$, $\la 1-q \ra$,
can be fitted in a similar way.

We also tried the converse operation, namely a scaling plot of
the data for periodic bc but using the value $\mu=0.63$. We find 
that one can get a relatively good data collapse excluding the
sizes $L=3,4$, and 5, which deviate from the scaling curve by less than
$10\%$. However, now the data for a given $\epsilon$ 
approach the scaling curve from the right side instead of from the left side
as in  Fig.~\ref{scalingratioql42}, but since they have an upward curvature, 
the correction to scaling should change curvature twice as $L$ increases,
which is not very plausible.
Hence, we believe that it is more natural to conclude that the
correct value of $\mu$ is closer to 0.42 than to 0.63, namely
that corrections to scaling are smaller for periodic than for free.

\begin{figure}
\begin{center}
\myscalebox{\includegraphics{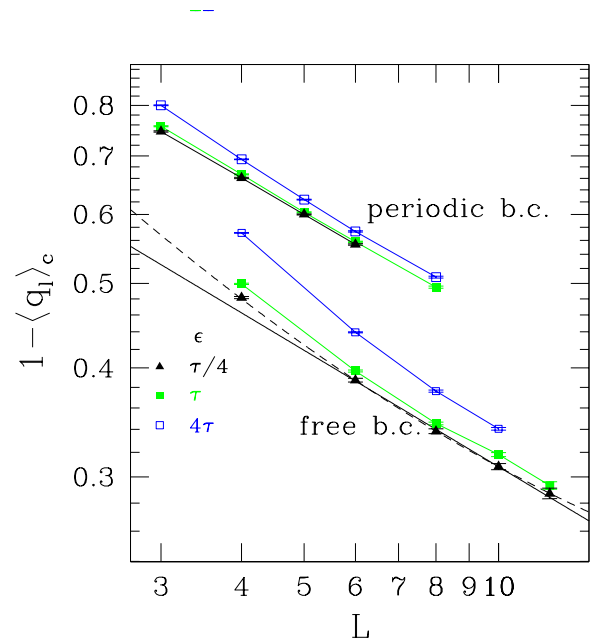}}
\caption{This plot shows together the data of Fig.~\ref{qllim} for
free boundary conditions and Fig.~\ref{qllimper} for periodic boundary 
conditions.}
\label{qlperiodicfree}
\end{center}
\end{figure}

Indeed, in general, it is reasonable to expect that 
corrections are larger for free bc, because these bc have a 
free surface on which lie a fraction of
sites which is quite substantial for moderate sizes. 
In Fig.~\ref{qlperiodicfree} 
we plot together the $\la 1-\ql \ra_c$ data of Fig.~\ref{qllim} and 
\ref{qllimper} for free and periodic bc.
The data for free bc lie significantly below those for periodic bc,
indicating that the surface of the excitations is smaller for free bc.
For periodic bc, the domain
wall has to ``bend'' to return to the same
point on the ``top surface'' as it
had on the ``bottom surface''. This may be the source of the extra surface
area.  
 
Under the hypothesis  $d-d_s \simeq 0.44$, Fig.~\ref{qlperiodicfree} then shows
 that the corrections to {\em asymptotic}\/ scaling
are larger for free bc, since the free bc data  show a more marked deviation
from the asymptotic $\epsilon$-independent behavior, and display a larger
curvature.
This is further indication that free bc have larger corrections.

Evidence that free bc have larger corrections was also
found recently in Monte Carlo simulations\cite{helmut}, where 
some evidence was  observed that the free bc data might have
a crossover from droplet-like to either TNT- or RSB-like behavior at 
large sizes. 

Incidentally, note that if RSB is the correct asymptotic picture
{\em and} the $L\to\infty$
limit of $\la 1-\ql \ra_c$ is the same for periodic and free bc, then 
Fig.~\ref{qlperiodicfree} would indicate that the corrections
are {\em smaller}\/ for free bc (since the data are closer to their
non-zero asymptotic value) which is not very plausible. 
Note, however, that we do not have an argument why in
the thermodynamic limit $\la 1-\ql \ra_c$ should be independent 
of boundary conditions.


\vspace{0.5cm}

To summarize the first part of the article, we have analyzed 
several quantities for periodic and free bc.
For both types of boundary conditions, 
all the data are well described by a general scaling 
picture involving only two scaling exponents, $d-d_s$ and $\theta'$, 
with only small corrections to scaling. Some observables
show significant corrections to {\em asymptotic}\/ scaling,
which are larger for free boundary conditions.
Fitting this scaling picture to our data, we obtain comparable 
values of $d-d_s$ for periodic ($0.43 \pm 0.02$) and free boundary
conditions ($0.44 \pm 0.03$).

By imposing that corrections to scaling are less than the statistical
errors of $1\%$, for periodic boundary conditions we obtain 
$\theta' \simeq 0$, which 
fits well the  TNT scenario ($d-d_s > 0, \, \theta' = 0$), while for 
free boundary conditions we obtain $\theta'=0.19 \pm 0.06$, 
which fits well the droplet picture  ($d-d_s > 0, \, \theta' > 0$).
By relaxing this requirement and allowing larger corrections to scaling
of order 10$\%$,
the data for free bc can be also fitted by a scenario with $\theta' \simeq 0$.
Therefore the data for free bc are  also consistent
with the TNT picture provided moderate corrections to scaling are allowed,
larger than those for periodic bc. We have also provided direct evidence that indeed
free bc have larger corrections
to scaling.  

Data for the box overlap for free boundary conditions 
indicates smaller corrections to asymptotic scaling, which is reasonable 
since the box is away from the surface, and are consistent
with the scenario described above.

For both free and periodic bc, the data are also fitted well by
the RSB picture  ($d-d_s = 0, \, \theta' = 0$), but only if we allow very 
large corrections to scaling. In this case, the good 
scaling behavior we observed for all the observable considered
would only be a finite size artifact, and would disappear at larger 
sizes. To test this possibility, large system sizes will be needed.

This concludes the first part of the paper, dedicated to the physical 
results.
In the second part, we will describe the branch and cut algorithm
employed, and analyze its performance in our computations.

\section{Branch and cut algorithm}
\label{sec:b&c}

\newcommand{\R}{{\mathbb{R}}}

Branch and cut is, to our knowledge, the 
fastest exact method 
for determining ground states of spin glasses in three dimensions.
To apply this technique, we transform the problem of minimizing the
Hamiltonian in Eq.(\ref{eq:ham}) into a standard combinatorial
optimization problem known as the {\em maximum cut} problem. (For a
detailed description of optimization and related topics, see
Ref.~\onlinecite{handbook}.) Consider the interaction graph $G=(V, E)$
associated with the spin glass Hamiltonian, where $G$ contains vertices $1,
\ldots, L^{3} \in V$ associated with the spin sites and edges $(ij) \in
E$ with weight $c_{ij} = -J_{ij}$ associated with the couplings.

Given a partition of $V$ into two sets, $W\subset V$ and its
complement $V \setminus W$, the {\em cut}\/ $\delta(W)$ associated
with $W$ is defined as the set of edges with one endpoint, $i$ say, in
$W$ and the other endpoint, $j$ say, in $V \setminus W$. In formulas,
$\delta(W) = \{(ij) \in E \mid i \in W, j \in V \setminus W \}$. The
{\em weight}\/ of a cut $\delta(W)$ is defined as the sum of the weights of
the cut edges $\sum_{(ij) \in \delta(W)} c_{ij}$. A {\em maximum cut}\/
is a cut with maximum weight among all partitions $W$.
It  is easy to show that minimizing the Hamiltonian
Eq.(\ref{eq:ham}) is equivalent to finding a maximum cut in $G$, see
Ref.~\onlinecite{desim}.
If we know a maximum cut with node partition $W$ and $V \setminus W$,
the corresponding ground state spin configuration can be read off by
assigning the value up to the spins in $W$ and down to the spins in $V
\setminus W$, or vice versa.

The branch-and-cut algorithm solves the maximum cut problem through
simultaneous lower and upper bound computations. By definition, the
weight of any cut gives a {\em lower bound}\/ on the optimal cut
value. Thus, we can start from any cut and iteratively improve the
lower bound using deterministic heuristic rules (local search and other specialized
heuristics, see Ref.\onlinecite{juen-vlsi} for details).
How do we decide when a cut is optimal? This can be done by
additionally maintaining {\em upper bounds}\/ on the value of the
maximum cut. Upon iteration of the algorithm, progressively tighter
bounds are found, until optimality is reached. 

Since the availability of upper bounds marks the difference between a
heuristic and an exact solution, we now summarize how the upper bound
is computed (for more details, see Ref.~\onlinecite{juen-vlsi}.)
To each edge $(ij)$ we associate a real variable $x_{ij}$ and to each
cut $\delta(W)$ an {\em incidence vector} $\chi^{\delta(W)} \in
\R^{E}$ with components $\chi_{ij}^{\delta(W)}$ associated to each
edge $(ij)$, where $\chi_{ij}^{\delta(W)}=1$ if $(ij)\in \delta(W)$
and $\chi_{ij}^{\delta(W)}=0$ otherwise. Denoting by $P_C(G)$ the
convex hull of the incidence vectors, it can be shown that a basic
optimum solution\cite{chvatal} of the linear program
\begin{equation}
  \label{eq:mc-formulation}
\max \{ \sum_{(ij) \in E} J_{ij}x_{ij} \mid x \in P_{C}(G)\}.  
\end{equation}
is a maximum cut. In order to solve (\ref{eq:mc-formulation}) with
linear programming techniques we would have to express $P_{C}(G)$ in
the form
\begin{equation} 
  P_{C}(G) = \{x \in \R^{E} \mid Ax \leq b, 0 \leq x \leq 1\}
\end{equation}
for some matrix $A$ and some vector $b$. Whereas the existence of $A$
and $b$ are theoretically guaranteed, even subsets of $Ax \leq b$
known in the literature contain a huge number of inequalities that
render a direct solution of (\ref{eq:mc-formulation}) impractical.

Instead, the branch-and-cut algorithm proceeds by optimizing over a
{\em superset}
$P$ containing $P_{C}(G)$, and by iteratively tightening $P$,
generating in this way progressively better upper bounds. The 
supersets
$P$ are generated by a {\em cutting plane}\/
approach. Starting with some $P$, we solve the linear program
$\max \{ \sum_{(ij) \in E} J_{ij} x_{ij} \mid x \in P \}$
by Dantzig's simplex algorithm\cite{chvatal}. Optimality is proven
if either of two conditions is satisfied: (i) the optimal value equals
the lower bound; (ii) the solution vector $\bar{x}$ is the incidence
vector of a cut.

If neither is satisfied, we have to tighten $P$ by solving the {\em
  separation problem}. This consists in identifying inequalities that
are valid for all points in $P_{C}(G)$, yet are violated by $\bar{x}$,
or reporting that no such inequality exists. The inequalities found in
this way are added to the linear programming formulation, obtaining a
new tighter partial system $P'\subset P$ which does not contain
$\bar{x}$. The procedure is then repeated on $P'$ and so on.

At some point, it may happen that (i) and (ii) are not satisfied, yet
the separation routines do not find any new cutting plane. In this
case, we {\em branch} on some fractional edge variable $x_{ij}$ (i.e.
a variable $x_{ij} \not \in \{0,1\}$), creating two subproblems in
which $x_{ij}$ is set to 0 and 1, respectively. We then we apply the
cutting plane algorithm recursively for both subproblems.

\section{Performance of the branch and cut algorithm}
\label{sec:performance}

In this section we study the performance of  our current
implementation of the branch and cut algorithm,
in particular the dependence of the number of computer operations on system
size. The results for size $L=12$ were obtained with a more
efficient version of the code, so performance for this size cannot be
compared with that for the smaller sizes. Hence, in this section,
we shall just consider sizes up to $L=10$.

Finding the ground state of the Hamiltonian Eq. (\ref{eq:ham}) in three
dimensions is an $\mathcal{NP}$-hard problem\cite{barahona}, 
and all known algorithms
to solve this class of problems require a number of operations that grows
exponentially on the size of the input, in the worst case.

However, depending on the problem, the number of operations for {\em
typical} instances (for the spin glass problem,
an instance is a realization
of the random couplings, or sample), can grow considerably more slowly 
than the worst case exponential behavior. Furthermore, the number of
operations can vary significantly from one instance to another.
It is therefore useful to investigate experimentally the performance
of the algorithm for typical instances, in order to try to extrapolate
the computational resources necessary to go to larger sizes, and
possibly to identify which parameters of the problem affect most the
performance.
De Simone et al.\cite{desim} measured the average CPU time used by the
branch and cut algorithm to find the ground state of the
two-dimensional $\pm J$ spin glass with periodic bc, up to $L=70$,
showing that the average CPU time was
approximated by a function proportional to $L^{6}$.

Here we analyze the performance of the branch and cut algorithm for
the three-dimensional spin glass with free bc and Gaussian
couplings. In order to do this, we first need a good measure of the
performance. For a complex algorithm such as branch and cut, a simple
and absolute measure of the number of operations is not available. Two
possible measures are the CPU time and the number of linear programs
solved during the run of the algorithm.
In Table \ref{tab_cpu}, we summarize the average running time needed for
calculating an unperturbed ground state for the different system
sizes.

\begin{figure}
\begin{center}
\myscalebox{\includegraphics{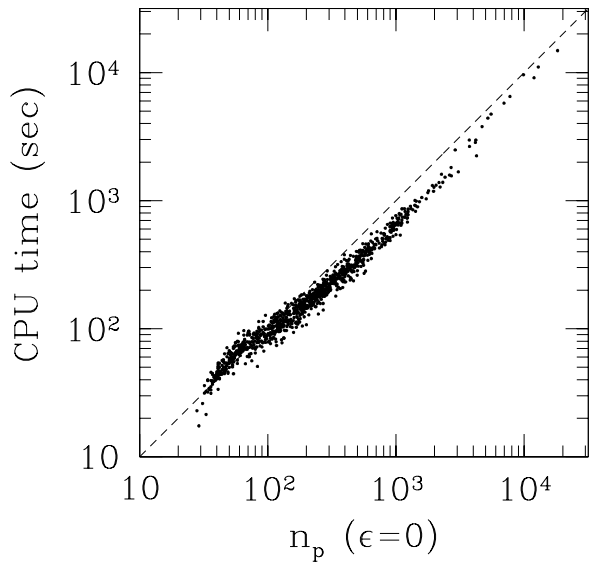}}
\caption{Scatter plot of the CPU time to find the unperturbed ground
state ($\epsilon=0$) versus the corresponding number of linear
programs solved ($n_p$).  Each point represents a randomly generated
sample with $L=10$. All the computations for this set of samples
were run on the same machine. The dashed line indicates a linear
behavior.}.
\label{scatter_lps_time}
\end{center}
\end{figure}

The CPU time is not an accurate measure since it depends on the 
machine architecture and load. Furthermore, our computations were carried out 
on several different machines, so the CPU time is not useful here.
We take instead the 
number of linear programs solved, $n_p$,
because (i) it is a well-defined and machine
independent quantity; (ii) we have observed that about $95\%$ of the time
is spent in solving linear programs; (iii) for a fixed system size,
$n_p$ correlates strongly, and almost linearly, with the CPU
time. This is shown in Fig.~\ref{scatter_lps_time}, which plots
the CPU time versus $n_p$ for 1000 randomly
generated samples with $L=10$, computed on the same machine.  
Note that since the {\em size} of the linear programs is also growing 
with the system size, the CPU time per linear program increases
strongly with $L$: 
the average (resp. median) CPU time goes from 
0.00770 (resp. 0.833) seconds for $L=4$ to 0.833 (resp. 0.784) seconds for
$L=10$.

Hence, $n_p$ severely underestimates the  rate at which  the
number of operations increases with $L$.

\begin{table}
\begin{center}
\begin{tabular}{r@{\hspace{0.9cm}}r}
\hline
\hline
$L$ & mean CPU time per sample\\
\hline
4   & $   0.065$ 
\\
6   & $   0.662$ 
\\ 
8   & $  10.11$  
\\
10  & $ 338 $    
\\
\hline
\hline
\end{tabular}
\end{center}
\caption{ 
  Mean CPU time per sample in seconds for the calculation of the unperturbed ground 
state, averaged also over the different machines.}
\label{tab_cpu}
\end{table}

\begin{figure}
\begin{center}
\myscalebox{\includegraphics{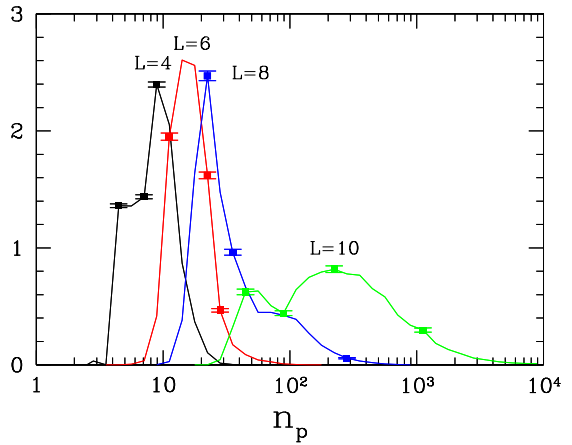}}
\caption{Histogram of the number of linear programs solved
by the branch and cut algorithm to find the unperturbed ground state
for different system sizes.}
\label{histo_lps}
\end{center}
\end{figure}

\begin{figure}
\begin{center}
\myscalebox{\includegraphics{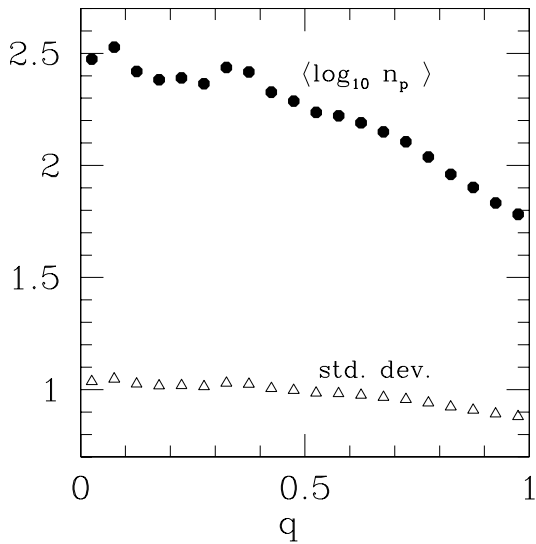}}
\caption{
The circles are a plot of $\langle \log_{10} n_p \rangle$,
where $n_p$ is the
number of linear programs solved
to compute the unperturbed ground state $S^{0}$, versus the overlap
between $S^{(0)}$ and the perturbed ground state $\tilde{S}^{(0)}$.
The data is for
$\epsilon/\tau=4$ and the samples were selected from a set of
randomly generated samples with $L=10$, in such a way that the same
number of samples is plotted for each consecutive $q$ interval of
length 0.1, in order to sample equally all regions of $q$. 
The triangles show the standard deviation, among samples, of $\log_{10} n_p$ as
a function of $q$.
}
\label{scatter_lps_q}
\end{center}
\end{figure}

From Fig.~\ref{scatter_lps_time}, we also note that the distribution
of $n_p$ (and CPU times) is  very broad, extending over three orders
of magnitude. The histogram distribution of $n_p$ for different system sizes
is shown in Fig.~\ref{histo_lps}. 
In addition to shifting to larger
$n_p$, the distribution broadens as $L$ increases. Also, there is some
evidence of a double-peak structure. For $L=10$, we verified that the
peak at smaller $n_{p}$ corresponds to samples
that could be solved without branching, while the other peak corresponds to 
samples where branching was necessary. Since in each branching step
the number of subproblems to be solved doubles, the number of linear
programs increases rapidly and the second peak is at significantly
larger $n_p$.

In order to identify which parameters of the problem, in addition to
the size, affect the performance, we ask
whether $n_p$ correlates with the physical observables we measure.
No significant correlation was observed with the 
ground state energy. Fig.~\ref{scatter_lps_q} 
plots\cite{note_scatter} $\langle \log_{10} n_p \rangle$
for the unperturbed ground state ($\epsilon=0$) and $L=10$
versus the overlap between this state and the perturbed
ground state with $\epsilon/\tau=4$.
We observe a distinct correlation between $n_p$ and $q$: for small $q$, more
linear programs are needed than for large $q$. The figure shows that the
typical number of linear programs is close to an order of magnitude larger if
$q \simeq 0$ than if $q \simeq 1$.
We observed a similar correlation 
for other values of $\epsilon$ as well, and also 
between the CPU time and $q$. Again, 
the distribution of $n_p$ is quite broad as
shown by the data for the standard deviation of
$\log_{10} n_p$ in Fig.~\ref{scatter_lps_q}.

\begin{figure}
\begin{center}
\myscalebox{\includegraphics{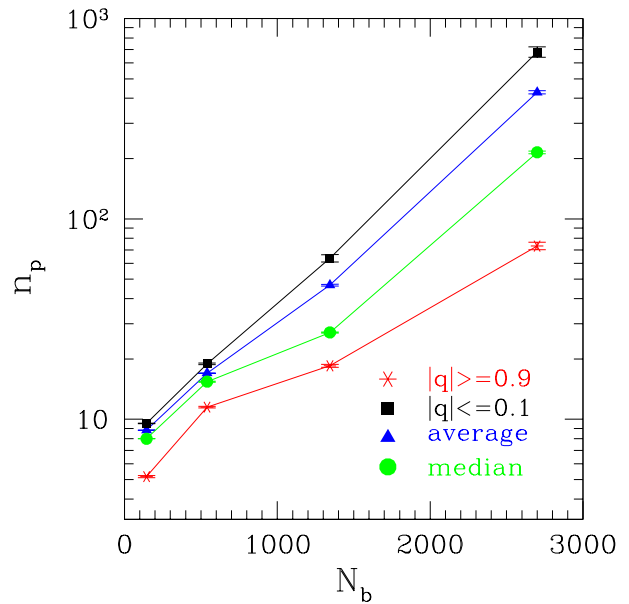}}
\caption{
Average $n_p$, median $n_p$, 
and conditional averages of $n_p$
restricted to $|q|\le 0.1$ and to $|q|\ge 0.9$, as a function of
the number of bonds $N_b$. The data for
$n_p$ are for $L=10$ and $\epsilon=0$ (unperturbed ground state),
and $q$ is the overlap between the $\epsilon=0$ and $\epsilon/\tau=4$
ground states.}
\label{lps}
\end{center}
\end{figure}

In order to quantify how the correlation between $n_p$ and $q$ changes
with the system size, we show in Fig.~\ref{lps} the average and median
of $n_p$ as a function of $N_b$, as well as the
conditional averages of $n_p$ restricted to samples with large
($|q|\ge 0.9$) and small ($|q|\le 0.1$) overlap.
We take the number of bonds, $N_b$, as a measure of the input size,
since the maximum cut problem is specified in terms of the edge
variables in the graph.
From Fig.~\ref{lps} we see that, first, all measures show an approximately 
exponential increase with $N_b$, with
corrections for small $N_b$, and  second, the difference between the
conditional averages with small and large $q$ seems to increase 
with the system size, and is about one order of magnitude for $L=10$.

A qualitative difference between samples with small and large overlap is
that samples with a small $|q|$ have a  rougher ``energy
landscape'', namely states with an energy close to the ground
state energy yet a spin configuration very different from the ground state. 
It is then intuitively clear why one would observe a correlation 
between $q$ and the running time for a 
stochastic algorithm employing local search heuristics, such as
simulated annealing, since when the algorithm encounters one of 
these configurations with small overlap, it must retrace its steps by a large amount. 

For the branch and cut algorithm,
the reason for the correlation between
$n_p$ and $q$ is less obvious, but some insight is provided by an
analysis of ``reduced cost fixing''. 
This is a feature of the branch
and cut algorithm speeding up the computations.
In every iteration of the algorithm, reduced
cost fixing gives us a sufficient condition 
to decide which variables
(corresponding to the edges in the graph) have already attained their
optimal value. Therefore, we can
fix  the variables with  ``optimal'' status 
to their current value for all
the subsequent iterations of the algorithm, resulting in less
overall computational effort. The more variables that can be fixed,
the faster the algorithm is in practice.
 
Since the samples with small overlap have ``almost optimal'' solutions
with  spin configurations very different from the ground state, a smaller number
of variables can be fixed. Here we do not have the ``correct'' edge
values available until the end. 
As an example, we checked that for $L=10$ and $\epsilon=\tau$, for 100
randomly chosen samples with small overlap ($|q| \leq 0.1$), 
in average $409 \pm
39$ of the 2700 edge variables could be fixed in the first sub
problem, i.e. before branching takes place. In contrast, for 100
randomly chosen samples with big overlap ($|q| \geq 0.9$), $921\pm 34$
of the edge variables could be fixed in the first sub problem,
about twice as many.
Of course, the less variables that can be fixed in the first sub problem,
the more overall branching is necessary,
resulting in more overall computational effort for samples with small
overlap.

A consequence of the the broad distribution of the CPU time and of its
correlation with the physical observables of interest, is that
a cutoff in the CPU time produces a systematic error in
these quantities.  One has therefore to ensure that the cutoff is
large enough so that the systematic error is smaller than the
statistical error. 

It is interesting to try to extrapolate the 
running time needed to deal with larger sizes.
The average CPU time
in Table \ref{tab_cpu} varies approximately as $\sim \exp(\alpha N_b)$ 
with $\alpha$ somewhere between 0.0024 and 0.003. 
Extrapolating to $L=14$
($N_b=7644$), this gives an average CPU time of around $10^{8\pm 1}$
seconds per sample, which is clearly very demanding.
Furthermore, memory limitations will set in before we can reach this
size. Again, note that $n_{p}$ increases much more slowly with
$N_b$. The data for $|q|\le 0.1$ in Fig.\ref{lps}, for example, 
vary approximately as $\sim \exp(\alpha N_b)$  with a smaller
$\alpha$ around
0.0017, showing that the dominant limiting factor is the solution 
of the linear programs.  
Note that the program used for $L=12$ is significantly faster 
than that used in this extrapolation.
This long extrapolated running time gives us further
motivation to continue our research on the improvement of this
algorithm.

\section{Conclusions}
\label{sec:discussion}
Using an \textit{exact} ``branch and cut''
optimization algorithm,
we have studied the large-scale, low-energy excitations in the Ising spin
glass in three dimensions with \textit{free} boundary conditions, and compared
the results with those obtained earlier by PY
for periodic boundary conditions.

In the first part of the paper, we have discussed in detail how the
whole set of observables analyzed is fitted by a general 
scaling picture characterized by two exponents, $d-d_s$ and $\theta'$,
and how the values of these parameters predicted by the various 
physical pictures proposed for the spin glass phase fit our data.
Our conclusions have been summarized at the end of
 Section \ref{subsec:discussion}.

In the second part of the paper, 
we have analyzed the performance of the branch and cut 
algorithm,  finding that the
performance is worse when there is a low energy excited state close in energy
to the ground state but far away in configuration space, and have given a
quantitative analysis of this effect.

\acknowledgements
We would like to thank A.~J.~Bray, G.~Parisi, M.~M\'ezard, D.~S.~Fisher,
and M.~A.~Moore  for helpful discussions and correspondence. APY acknowledges
support from the NSF through grant DMR 0086287. MP would
like to thank A.~J.~Bray for a useful suggestion on the data analysis.
We thank the Regional Centre of Computing of the University of Cologne
for the allocation of computer time. Over the years, Giovanni Rinaldi
and Gerd Reinelt contributed much to the algorithm.


\begin{thebibliography}{}
\bibitem{reality}
   In reality, our implementation of the branch and cut algorithm works
 as follows: If we have found the optimal configuration, i.e. two sets
 of spins where all spins in one set point up and all spins in the
 other point down, we assign the value ``up'' to the set with the
 smaller cardinality. Therefore we do not produce a symmetrical
 distribution. This does not affect the analysis since we are interested
 only in the absolute value of the overlap.
\bibitem{interval}
    This average does not coincide with the midpoint of the interval because
    the perturbation method does not sample uniformly in $q$.
\bibitem{errors}
   {\bf changed} The statistical errors on the fit parameters are given
    by the value of the parameter that increases the $\chi^2$ by one
  with  respect to the best fit (lowest $\chi^2$), when the other parameters 
  are kept fixed.     Hence they do not take into account
    correlations between the various fit parameters, and can underestimate
    the actual statistical uncertainty.
    Furthermore, the range of acceptable
    parameters is usually larger than the condition $\Delta \chi^2=1$, 
  since larger values of $\chi^2$ can still be statistically acceptable.
\bibitem{note_scatter}
The perturbation method does not
generate a uniform distribution of $q$, therefore
Fig.~\ref{scatter_lps_q} was produced by selecting 1000 samples from a
random ensemble, such that there is the same number of samples in each
consecutive $q$ interval of length 0.1 in the range $q\in [-1,1]$.
\end{thebibliography}
\end{document}